\documentclass[%
reprint,
superscriptaddress,
amsmath,
amssymb,
aps,
prl,
floatfix,
]{revtex4-2}

\usepackage{float}
\usepackage{xcolor}
\usepackage{graphicx}% Include figure files
\usepackage{dcolumn}% Align table columns on decimal point
\usepackage{bm}% bold math
\usepackage{comment}
\usepackage{soul}
\usepackage{siunitx}
\usepackage{mathtools}

\usepackage[unicode=true,pdfusetitle,bookmarks=false,colorlinks=true,citecolor=blue,urlcolor=blue,linkcolor=blue]{hyperref}

\begin{document}

\preprint{APS/123-QED}

\title{
Structure and Memory Control Self-Diffusion in Active Matter
%Structure and Memory Control Self-Diffusion in Active Brownian Liquids
%Structure and Memory Control Self-Diffusion in Active Matter
%Self-Diffusion in Interacting Active Brownian Particles
}% Force line breaks with \\
%\thanks{A footnote to the article title}%

\author{Akinlade Akintunde}
\affiliation{Department of Chemistry, The Pennsylvania State University, University Park, Pennsylvania, 16802, USA}

\author{Stewart A. Mallory}
\email{sam7808@psu.edu}
\affiliation{Department of Chemistry, The Pennsylvania State University, University Park, Pennsylvania, 16802, USA}
\affiliation{Department of Chemical Engineering, The Pennsylvania State University, University Park, Pennsylvania, 16802, USA}

%\collaboration{CLEO Collaboration}%\noaffiliation

\date{\today}% It is always \today, today,
             %  but any date may be explicitly specified

\begin{abstract}
Despite extensive progress in characterizing the emergent behavior of active matter, the microscopic origins of self-diffusion in interacting active systems remain poorly understood.
Here, we develop a framework that quantitatively links self-diffusion to collisional forces and their temporal correlations in active fluids.
We show that transport is governed by two contributions: an equal-time suppression of motion arising from anisotropic collisional forces, and a memory correction associated with the temporal persistence of these forces.
Together, these effects yield an exact expression for the self-diffusivity in terms of measurable force statistics and correlation times.
We apply this framework to purely repulsive active Brownian particles and find that self-diffusion is always reduced, with collisional memory acting as a strictly dissipative correction.
Our results establish a direct connection between microscopic force correlations and macroscopic transport, providing a general mechanical perspective for interpreting self-diffusion in active matter.
\end{abstract}

\maketitle

\section{Introduction}

Self-diffusion provides a direct window into how microscopic forces and fluctuations generate macroscopic transport in liquids~\cite{dhont1996}.
Together with the mean-squared displacement (MSD), these quantities serve as a probe of local structure, force correlations, and dynamical constraints.
In equilibrium systems, diffusion is closely tied to dissipation through relations such as Stokes--Einstein, while deviations reveal hydrodynamic correlations and many-body effects~\cite{chapman1990mathematical,Zahn1997-ne,zwanzig2001nonequilibrium,Bian2016-xu,balakrishnan2020elements}.
In dense colloidal suspensions, the suppression of self-diffusion signals the onset of caging and dynamical arrest~\cite{Weeks2002-lo,Berthier2011-sc,Hunter2012-qf,Berthier2014-la}.
In general, self-diffusion and the MSD act as sensitive diagnostics of the mechanisms governing transport.

Active fluids share many structural features with equilibrium liquids but are driven far from equilibrium by persistent self-propulsion~\cite{Romanczuk2012-dw,Nardini2017-ni,Fodor2016-nb}.
Here, propulsion replaces thermal agitation as the dominant source of fluctuations~\cite{Paxton2004-vw,Golestanian2005-pf,Wang2006-je,Ghosh2009-ok,Ebbens2010-nb,Gao2012-gc,Izri2014-dh,Moran2017-yk}, fundamentally altering how particles explore their environment~\cite{Romanczuk2011-ym}.
While effective interactions, phase separation, and collective dynamics in active systems have been widely studied~\cite{schweitzer2003brownian,Nelson2010-jh,Sanchez2011-ej,Wang2012-kd,Soler2013-vk,Patra2013-by,Garcia-Gradilla2013-zk,Guix2014-jz,Gao2014-bq,Li2016-yv,Needleman2017-lh,Dulaney2021-kf,Choi2025-mu}, a unified microscopic framework for self-diffusion in interacting active liquids remains lacking.
Most existing theories focus on isolated particles or dilute suspensions~\cite{Ghosh2015-xp,Howse2007-vh,Sevilla2014-yt,Sevilla2015-jn,Basu2018-th,Dulaney2020-mb,Mallory2020-tq,Sandoval2020-yy,Schakenraad2020-dg,Caraglio2022-dz,Modica2023-tb,Bayati2024-kd,Bayati2025-kk,Soto2025-lg}, leaving the role of interparticle interactions across the liquid regime poorly understood.
Because self-diffusion is a macroscopic transport coefficient that sets the timescale over which active particles explore space, relax density fluctuations, and reorganize local structure, a microscopic description is essential for developing coarse-grained and continuum theories of active matter~\cite{julicher2018hydrodynamic,te2025metareview,ramaswamy2019active}.

Here, we develop a microscopic framework for self-diffusion in active Brownian particle (ABP) systems that explicitly incorporates interparticle interactions.
By exploiting a general steady-state property of ABPs, namely the orthogonality between particle velocities and conservative collisional forces~\cite{Schiltz-Rouse2023-vj,akintunde2025single,Omar2023-lq}, we derive an exact expression for the long-time self-diffusivity.
The theory reveals that diffusion is governed by two distinct but related contributions:
(i) an equal-time suppression of particle motion arising from anisotropic collisional forces aligned with the propulsion direction, and
(ii) a memory correction associated with the temporal persistence of these collisional correlations.
Together, these contributions connect long-time transport directly to measurable force statistics and correlation times.

We apply this framework to a suspension of purely repulsive ABPs, which we use here as a controlled and extensively studied reference system~\cite{Shaebani2020-yf}.
By decomposing the dynamics into a small set of tractable correlation functions, we quantify how activity and crowding regulate both the MSD and the long-time self-diffusivity across the liquid regime.
Particle-based simulations validate theoretical predictions and show that equal-time force fluctuations dominate the reduction in self-diffusivity, while collisional memory provides a systematic, strictly dissipative correction whose magnitude depends on activity and area fraction.

\section*{Model}

We consider a suspension of \(\mathcal{N}\) identical athermal active Brownian disks confined to a two-dimensional (2D) periodic square of area \(\mathcal{A}\), with number density \(\rho = \mathcal{N}/\mathcal{A}\).
While we focus on 2D for concreteness, the theoretical framework developed here carries over unchanged to 3D, with only a modification to the orientational dynamics.
Each particle is driven by a constant self-propelling force
\(\mathbf F_a = \gamma U_a \hat{\mathbf q}\)
along the orientation unit vector
\(\hat{\mathbf q}[\theta(t)] = \{\cos\theta(t), \sin\theta(t)\}\).
The orientation undergoes rotational Brownian motion characterized by a reorientation time \(\tau_R\).
Translational dynamics are overdamped with drag coefficient \(\gamma\).
Thermal noise is neglected, and interparticle interactions are modeled by a conservative collisional force \(\mathbf F_c\), assumed pairwise additive.
The equations of motion for a given particle are
\begin{subequations}
\label{eq:1}
\begin{align}
\gamma\,\dot{\mathbf r}
= \mathbf F_a + \mathbf F_c
&= \gamma U_a\,\hat{\mathbf q}\!\left[\theta(t)\right] + \mathbf F_c,\\
\dot{\theta}(t) &= \xi(t),
\end{align}
\end{subequations}
where \(\xi(t)\) is Gaussian white noise with zero mean,
\(\langle \xi(t) \rangle = 0\), and correlation
\(\langle \xi(t)\,\xi(t') \rangle = (2/\tau_R)\,\delta(t - t')\).
In the results presented below, we specialize to purely repulsive interparticle interactions; all other aspects of the theoretical framework remain general.

\section{Theory}

The mean-squared displacement (MSD) of a tagged particle can be written as
\begin{equation}
\label{eq:2}
\langle [\Delta\mathbf r(t)]^2 \rangle
=
2t
\int_0^t
\left(1-\frac{\tau}{t}\right)
\langle \mathbf v(\tau)\cdot\mathbf v(0) \rangle
\, d\tau \, ,
\end{equation}
where \(\mathbf v(t)=\dot{\mathbf r}(t)\) and
\(\langle \mathbf v(\tau)\cdot\mathbf v(0) \rangle\) is the velocity autocorrelation function (VACF)~\cite{dhont1996}.
In the long-time limit, Eq.~(\ref{eq:2}) yields the self-diffusivity,
\begin{equation}
\label{eq:3}
D
\equiv
\lim_{t\to\infty}
\frac{\langle [\Delta\mathbf r(t)]^2 \rangle}{4t}
=
\frac{1}{2}
\int_0^\infty
\langle \mathbf v(\tau)\cdot\mathbf v(0) \rangle
\, d\tau \, .
\end{equation}
Additionally, from Eq.~(\ref{eq:2}), it is straightforward to show that the particle motion at short times is ballistic and scales as
\(
\langle [\Delta\mathbf r(t)]^2 \rangle
\sim
\langle \mathbf v^2\rangle\, t^2
\).

Substituting the equations of motion [Eq.~(\ref{eq:1})] and introducing normalized force–force correlation functions
\(
C_{jk}(\tau)
=
 \langle \mathbf F_j(\tau)\!\cdot\!\mathbf F_k(0)\rangle
/
\langle \mathbf F_j(0)\!\cdot\!\mathbf F_k(0)\rangle
\),
the VACF can be written as
\begin{align}
\label{eq:4}
\gamma^2
&\langle \mathbf v(\tau)\!\cdot\!\mathbf v(0) \rangle
=
\langle \mathbf F_a^2\rangle \, C_{aa}(\tau)
\nonumber\\
&\quad+
\langle \mathbf F_a\!\cdot\!\mathbf F_c\rangle
\left[
C_{ac}(\tau)
+
C_{ca}(\tau)
\right]
+
\langle \mathbf F_c^2\rangle \, C_{cc}(\tau) \, .
\end{align}

As the propulsion force has a fixed magnitude, it follows that
\(\langle \mathbf F_a^2\rangle = (\gamma U_a)^2\).
A second simplifying relation follows from the condition that, in steady state, the particle velocity is statistically orthogonal to the collisional force~\cite{akintunde2025single,Schiltz-Rouse2023-vj,Omar2023-lq},
\begin{equation}
\label{eq:5}
\langle \mathbf v \!\cdot\! \mathbf F_c \rangle = 0 \, .
\end{equation}
This identity reflects the fact that conservative interparticle forces do no net work on average at steady state.
Using the equation of motion to eliminate the velocity in Eq.~(\ref{eq:5}) yields
\begin{equation}
\label{eq:6}
\langle \mathbf F_a \!\cdot\! \mathbf F_c\rangle
=
-\,\langle \mathbf F_c^2\rangle \, .
\end{equation}
As a consequence, the mean-square velocity can be written as
\begin{equation}
\label{eq:7}
\gamma^2 \langle \mathbf v^2\rangle
=
(\gamma U_a)^2
-
\langle \mathbf F_c^2\rangle \, .
\end{equation}
We refer to \(\langle \mathbf F_c^2 \rangle\)  as the collisional force variance, which quantifies the typical instantaneous force imbalance induced by interparticle interactions.
Applying these equal-time identities, the VACF simplifies to
\begin{align}
\label{eq:8}
\gamma^2\langle \mathbf v(\tau)\!& \cdot\!\mathbf v(0) \rangle
=
(\gamma U_a)^2 \, C_{aa}(\tau)
\nonumber\\
&\quad-
\langle \mathbf F_c^2\rangle
\left[
C_{ac}(\tau)
+
C_{ca}(\tau)
-
C_{cc}(\tau)
\right] \, .
\end{align}

The orientation autocorrelation function appearing in the first term of Eq.~(\ref{eq:8}) can be evaluated analytically, yielding
\begin{equation}
\label{eq:9}
C_{aa}(\tau)
=
\langle \hat{\mathbf q}(\tau) \cdot \hat{\mathbf q}(0) \rangle
=
e^{-\tau/\tau_R}.
\end{equation}
The two force-orientation correlations in the second term,
\(C_{ca}(\tau)\) and \(C_{ac}(\tau)\),
are not generally equal.
However, because the orientation evolves autonomously under rotational diffusion and is statistically independent of earlier collisional forces,
\(C_{ac}(\tau)=C_{aa}(\tau)\).
Using this identity, the VACF reduces to
\begin{align}
\label{eq:10}
\gamma^2  \langle \mathbf v(\tau)\! & \cdot\!\mathbf v(0) \rangle
 =
\nonumber\\
& \gamma^2 \langle \mathbf v^2\rangle \, e^{-\tau/\tau_R} -
\langle \mathbf F_c^2\rangle
\left[
C_{ca}(\tau)
-
C_{cc}(\tau)
\right] \, .
\end{align}
The exponential decay of the first term is set by rotational Brownian motion, with an amplitude \(\langle \mathbf v^2\rangle\) that incorporates interaction effects through the equal-time correlations in Eq.~(\ref{eq:7}).
The second term accounts for additional time-dependent contributions arising from correlations between collisional forces and the particle orientation, as captured by the difference \(C_{ca}(\tau)-C_{cc}(\tau)\).
Equation~(\ref{eq:10}) provides the most general form of the VACF consistent with steady-state ABP dynamics and autonomous rotational diffusion, with interaction effects entering only through the equal-time velocity variance and collisional memory correlations.

Substituting this expression [Eq.~(\ref{eq:10})] into the relation for the self-diffusivity [Eq.~(\ref{eq:3})] gives
\begin{equation}
\label{eq:11}
D
=
\frac{1}{2}
\langle \mathbf v^2\rangle \, \tau_R
-\frac{1}{2}
\frac{\langle \mathbf F_c^2\rangle}{\gamma^2}
\tau_D \, ,
\end{equation}
where
\begin{equation}
\label{eq:12}
\tau_D
=
\tau_{ca}
-
\tau_{cc}
=
\int_0^\infty
\left[
C_{ca}(\tau)
-
C_{cc}(\tau)
\right]
\, d\tau \, .
\end{equation}
The net collisional memory timescale \(\tau_D\) characterizes the persistence of collisional correlations contributing to velocity memory beyond orientational relaxation.
Using Eq.~(\ref{eq:7}), this result can be written as
\begin{equation}
\label{eq:13}
\frac{D}{D_0}
=
1
-
\frac{\langle \mathbf F_c^2\rangle}{(\gamma U_a)^2}
\left(
1
+
\frac{\tau_D}{\tau_R}
\right) \, ,
\end{equation}
where
\(D_0 = \frac{1}{2} U_a^2 \tau_R\)
is the ideal self-diffusivity in the absence of interparticle interactions
(\(\mathbf F_c = 0\)).
Equation~(\ref{eq:13}) makes explicit how deviations from the ideal active self-diffusivity arise from the collisional force variance \(\langle \mathbf F_c^2\rangle\), modulated by the relative importance of collisional memory through the factor \(1+\tau_D/\tau_R\).
While \(\langle \mathbf F_c^2\rangle\) is strictly positive, the sign of \(\tau_D\) is not fixed \textit{a priori}, since it originates from a difference of normalized force-force correlation functions.
As a result, interactions may either suppress or enhance long-time transport, depending on whether collisional memory opposes or reinforces persistent motion.
\begin{figure}[b!]
\centering
\includegraphics[width=.85\linewidth]{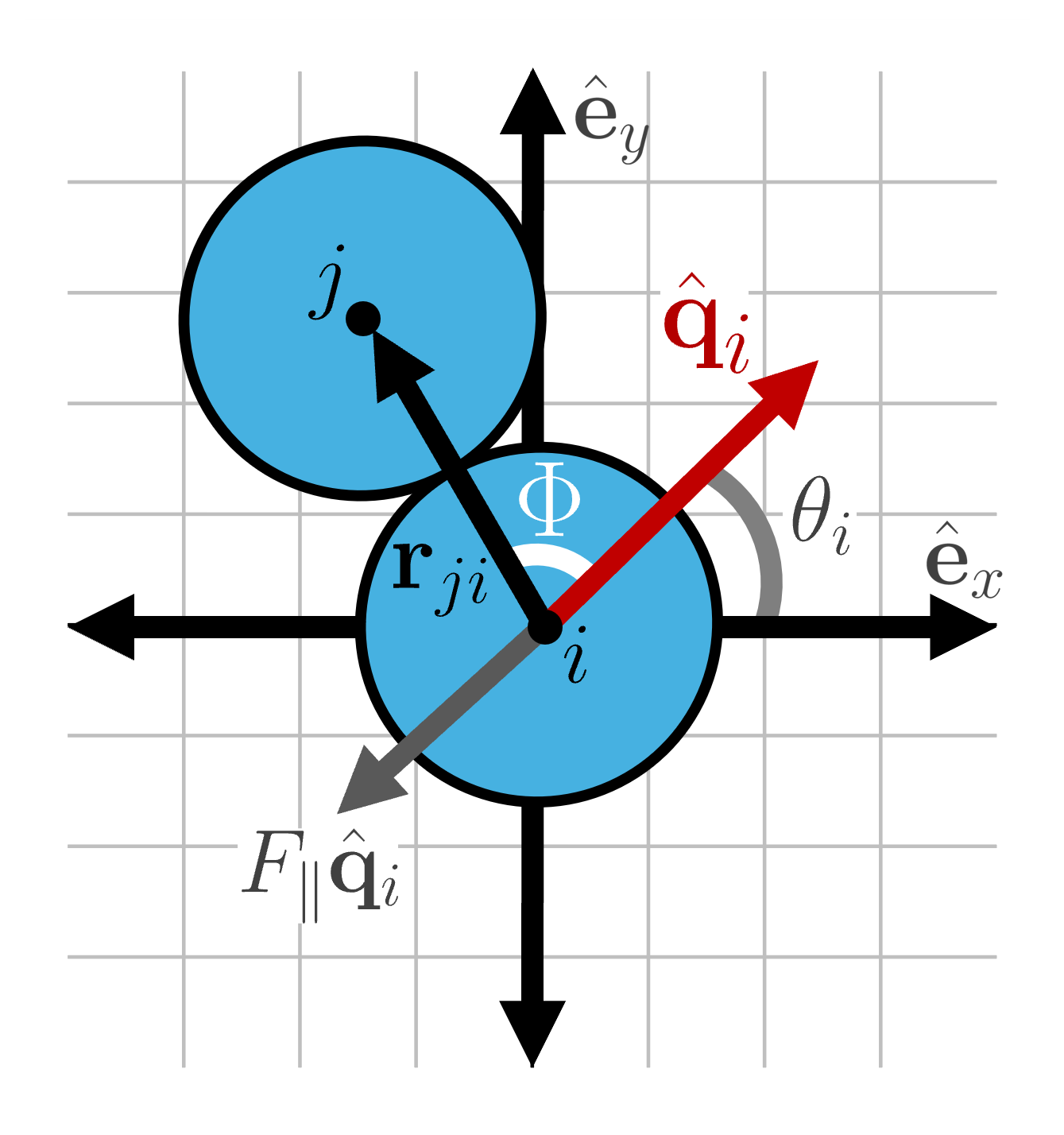}
\caption{
Schematic of the active Brownian particle model (ABP) and force geometry.
Particles self-propel along their orientation $\hat{\mathbf q}_i$ and interact via conservative forces acting along interparticle separation vectors $\mathbf r_{ji}$.
The projection of the collisional force onto the propulsion direction, \(F_{\parallel}\equiv \mathbf F_c\cdot \hat{\mathbf q}\), generates an anisotropic force imbalance that suppresses particle motion and underlies the reduction of self-diffusion.
}

\label{fig:1}
\end{figure}

Lastly, the collisional force variance \(\langle \mathbf F_c^2\rangle\), and hence the mean-square velocity \(\langle \mathbf v^2\rangle\) via Eq.~(\ref{eq:7}), can be related to the local structure around a particle by introducing the anisotropic pair correlation function~\cite{Bialke2013-gy,Speck2015-df,Wittkowski2017-ty,Jeggle2020-ca,Grossmann2020-th,Dasgupta2026-cz},
\begin{align}
\label{eq:14}
\langle \mathbf F_c^2\rangle
& =
-\langle \mathbf F_a \!\cdot\! \mathbf F_c\rangle  = - \gamma U_a \langle  |\mathbf F_c| \cos \Phi \rangle \nonumber \\
&=
\rho \gamma U_a
\int_{-\pi}^{\pi}
\int_0^{\infty}
V'(r)\,
g(r,\Phi)\,
r\cos\Phi
\,dr\,d\Phi
\, .
\end{align}
Here, \(\rho\) is the number density, \(V'(r)\) denotes the derivative of the interparticle pair potential, \(\Phi\) is the angle between the interaction vector \(\mathbf r_{ji}\) and the particle orientation \(\hat{\mathbf q}_i\) (see Fig.~\ref{fig:1}), and \(g(r,\Phi)\) is the anisotropic pair correlation function.

An instructive simplification follows by expanding the anisotropic pair correlation function \(g(r,\Phi)\) in angular harmonics about the propulsion direction.
In a homogeneous and isotropic steady state, symmetry implies that \(g(r,\Phi)\) is an even function of \(\Phi\), and therefore admits the Fourier expansion
\begin{equation}
\label{eq:15}
g(r,\Phi)
=
g_0(r)
+
\sum_{n=1}^{\infty}
g_n(r)\cos(n\Phi) \, ,
\end{equation}
where \(g_0(r)\) is the isotropic component of the pair correlation function and \(g_n(r)\) are the radial amplitudes of the higher angular modes.
Substituting Eq.~(\ref{eq:15}) into Eq.~(\ref{eq:14}) and using the orthogonality of trigonometric functions over \(\Phi\) yields
\begin{equation}
\label{eq:16}
\langle \mathbf F_c^2\rangle
=
\rho \gamma U_a\,\pi
\int_0^{\infty}
V'(r)\,
g_1(r)\,
r\,dr \, ,
\end{equation}
where
\begin{equation}
\label{eq:17}
g_1(r)
\equiv
\frac{1}{\pi}
\int_{-\pi}^{\pi}
g(r,\Phi)\cos\Phi\,d\Phi
\end{equation}
is the first angular moment of the pair correlation function.
Equation~(\ref{eq:16}) shows that the collisional force variance, and hence the reduction of the mean-square velocity, is controlled entirely by the dipolar component of the anisotropic pair structure relative to the propulsion direction.
While we focus on 2D systems for clarity, all aspects of the theory can be generalized to 3D.

\begin{figure*}[t!]
\centering
\includegraphics[width=0.95\linewidth]{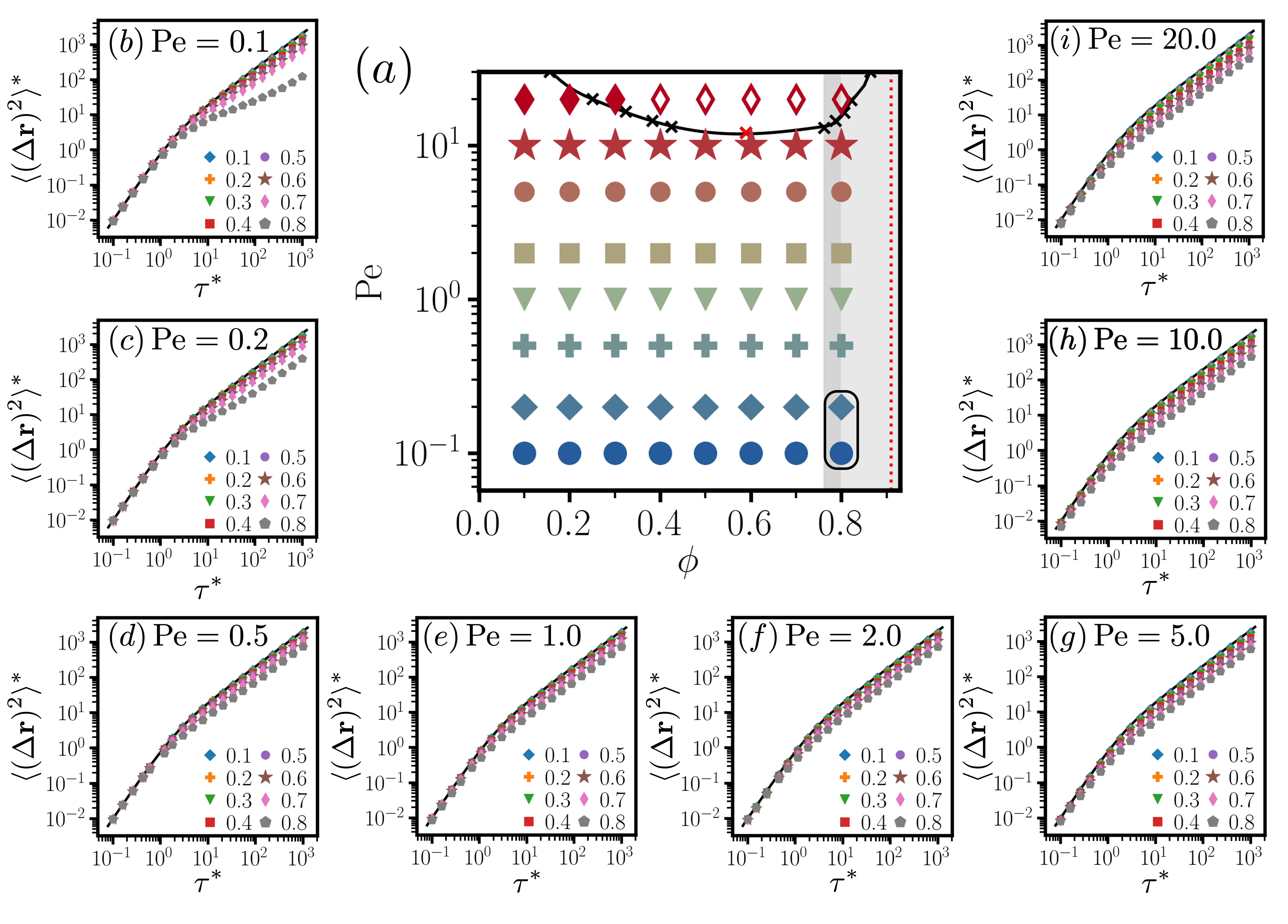}
\caption{
Simulation results for the 2D ABP system.
(a) Parameter space explored as a function of P\'eclet number $\mathrm{Pe}$ and area fraction $\phi$.
The solid black line denotes the binodal separating homogeneous states from the two-phase coexistence region for hard active Brownian disks~\cite{Levis2017-je}.
Light and dark gray regions indicate the hexatic and solid phases for passive hard disks, respectively~\cite{lee2016molecular}.
The red dotted line marks close packing in 2D, $\phi_{cp} \approx 0.91$.
State points enclosed by the black oval exhibit anomalous diffusion at intermediate times.
(b--i) Mean-squared displacement (MSD) for states at the indicated values of $\mathrm{Pe}$ and $\phi$.
The MSD and lag time are rescaled by $\langle \mathbf v^2\rangle \tau_R^2$ and $\tau_R$, respectively, collapsing the short-time ballistic regime.
The solid black line shows the MSD of an ideal 2D ABP.
}
\label{fig:2}
\end{figure*}

\section{Results}

In this section, we present Brownian dynamics simulation results for the MSD and the long-time self-diffusivity of purely repulsive ABPs (See Appendix for simulation details).
Our goal here is to establish the phenomenology of self-diffusivity in a canonical active system and to provide a concrete setting for applying the theoretical framework developed above.
A microscopic interpretation is deferred to the Discussion.

For ABPs, with hard or effectively hard repulsive interactions, the system state is determined by two parameters.
The Péclet number,
\(
\mathrm{Pe} = \ell_0 / \sigma
\),
quantifies activity through the intrinsic run length \(\ell_0 = U_a \tau_R\) relative to the particle diameter \(\sigma\).
The second parameter is the area fraction,
\(
\phi = \rho \mathcal{A}_p
\),
where \(\rho\) is the number density and \(\mathcal{A}_p = \pi\sigma^2/4\) is the area of a particle.

The phase behavior of purely repulsive active Brownian disks is well established~\cite{Fily2012-ej,Fily2014-vf,Levis2014-fe,Digregorio2018-ve,Mallory2021-hi}.
In the limit \(\mathrm{Pe} \to 0\), the system approaches the equilibrium hard-disk phase diagram, exhibiting a fluid--hexatic--solid sequence with increasing area fraction~\cite{Bernard2011-qm}.
As activity increases, the liquid regime broadens, and beyond a critical activity, the system undergoes motility-induced phase separation into dense and dilute phases.
In this study, we find that the onset of phase separation occurs at \(\mathrm{Pe}_c \approx 12\), consistent with prior studies.

Figure~\ref{fig:2}(a) summarizes the values of \(\mathrm{Pe}\) and \(\phi\) explored in this study and their location within the phase diagram.
We restrict attention to isotropic states below the onset of phase separation, focusing on \(\phi < 0.8\) and \(\mathrm{Pe} < \mathrm{Pe}_c\).
This choice allows for the examination of transport properties in well-defined liquid states.

Figures~\ref{fig:2}(b--i) show the MSD for each of the points in the representative \(\mathrm{Pe}-\phi\) state space.
In order to collapse the short-time ballistic regime, we nondimensionalize the MSD as,
\(
\langle (\Delta \mathbf r)^2\rangle^*
=
\langle (\Delta \mathbf r)^2\rangle / (\langle \mathbf v^2\rangle \tau_R^2)
\),
and plot it as a function of the dimensionless lag time \(\tau^* = t/\tau_R\).
The solid black line corresponds to the MSD of an ideal ABP in 2D~\cite{ten2011brownian}:   
\begin{equation*}
\langle (\Delta \mathbf r)^2\rangle^*
= 2 \Big[\tau^* + \big(e^{-\tau^*}-1\big)\Big]\, .
\end{equation*}
Across the parameter range studied, the MSD curves exhibit a high degree of self-similarity with the MSD transition from ballistic to diffusive after an approximate time \(\tau_R\).
Increasing area fraction systematically suppresses the MSD beyond the short-time ballistic regime for all values of \(\mathrm{Pe}\).
At intermediate times, deviations from simple ballistic-to-diffusive crossover emerge at large \(\phi\) and small \(\mathrm{Pe}\) [See Fig.~\ref{fig:2}(a)].

\begin{figure}[t!]
\centering
\includegraphics[width=0.85\linewidth]{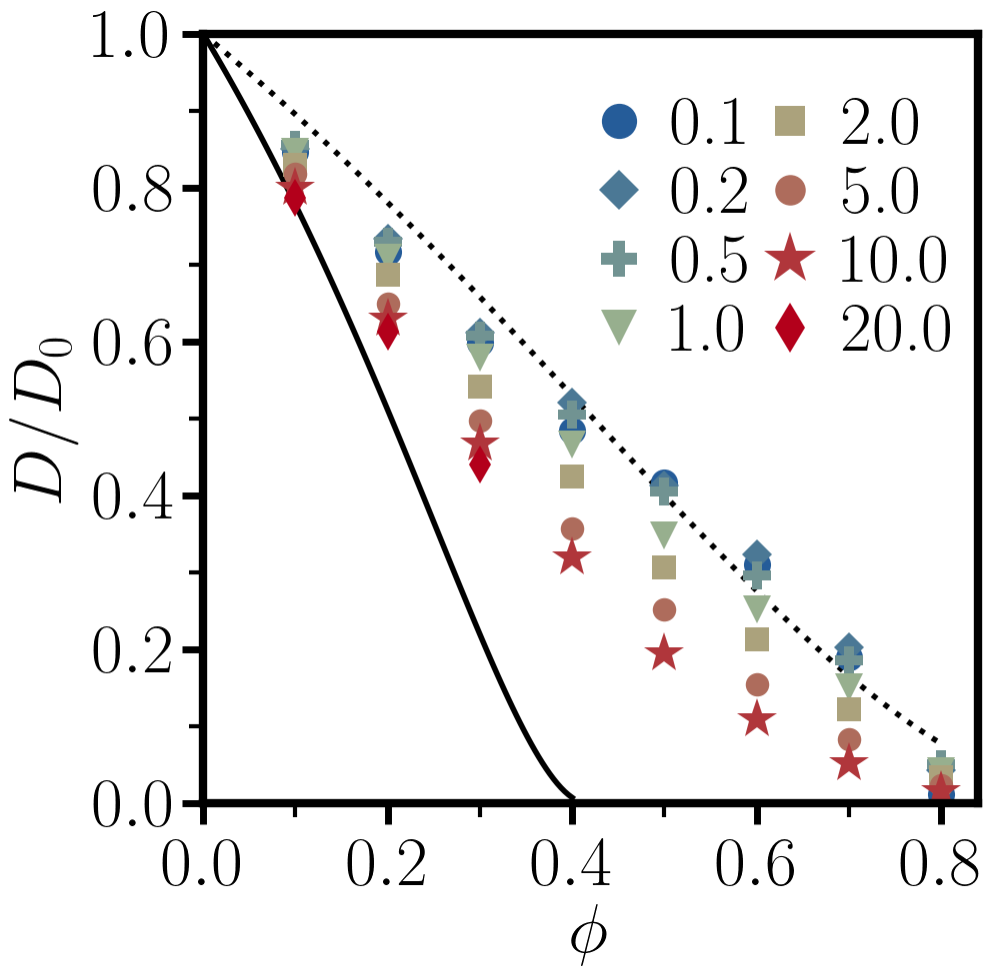}
\caption{
Normalized long-time self-diffusivity $D/D_0$ as a function of area fraction $\phi$ for various P\'eclet numbers.
Across the parameter range studied, increasing \(\phi \) leads to a monotonic suppression of self-diffusion.
The solid black line shows the analytical prediction for $D/D_0$ in the asymptotic limit $\mathrm{Pe}\to\infty$~\cite{Soto2025-lg}.
The dotted line corresponds to the analytical expression for passive Brownian disks~\cite{Stopper2018-nk}.
}
\label{fig:3}
\end{figure}

Figure~\ref{fig:3} reports the normalized self-diffusivity extracted from the MSD data.
In the joint limit \(\mathrm{Pe} \to 0\) and \(\phi \to 0\), interparticle interactions vanish, and ideal behavior is recovered.
For the representative \(\mathrm{Pe}-\phi\) state space studied, the normalized self-diffusivity decreases monotonically with increasing area fraction.
At fixed \(\phi\), increasing activity reduces \(D/D_0\) and leads to saturation at a finite value as \(\mathrm{Pe}\) approaches \(\mathrm{Pe}_c\).

In the equilibrium limit \(\mathrm{Pe} \to 0\), the normalized self-diffusivity approaches the analytical prediction for passive hard disks,
\(
D/D_0 = [1 + a\, g_{hd}]^{-1}
\),
where \(g_{hd} = (1 - \phi/2)/(1 - \phi)^2\) is the pair correlation function at contact~\cite{Stopper2018-nk} and \(a=1\).
This limiting behavior is shown as the dotted line in Fig.~\ref{fig:3}.
In the opposite limit of large activity, Soto \textit{et al.}~\cite{Soto2025-lg} predict
\(
D/D_0 = [1 - \phi\, g_{ed}]^2
\),
based on a density-dependent effective propulsion speed, where \(g_{ed} = (1 - 7\phi/16)/(1 - \phi)^2\).
This asymptotic prediction is shown as the solid line in Fig.~\ref{fig:3}.

\begin{figure*}
\centering
\includegraphics[width=0.9\linewidth]{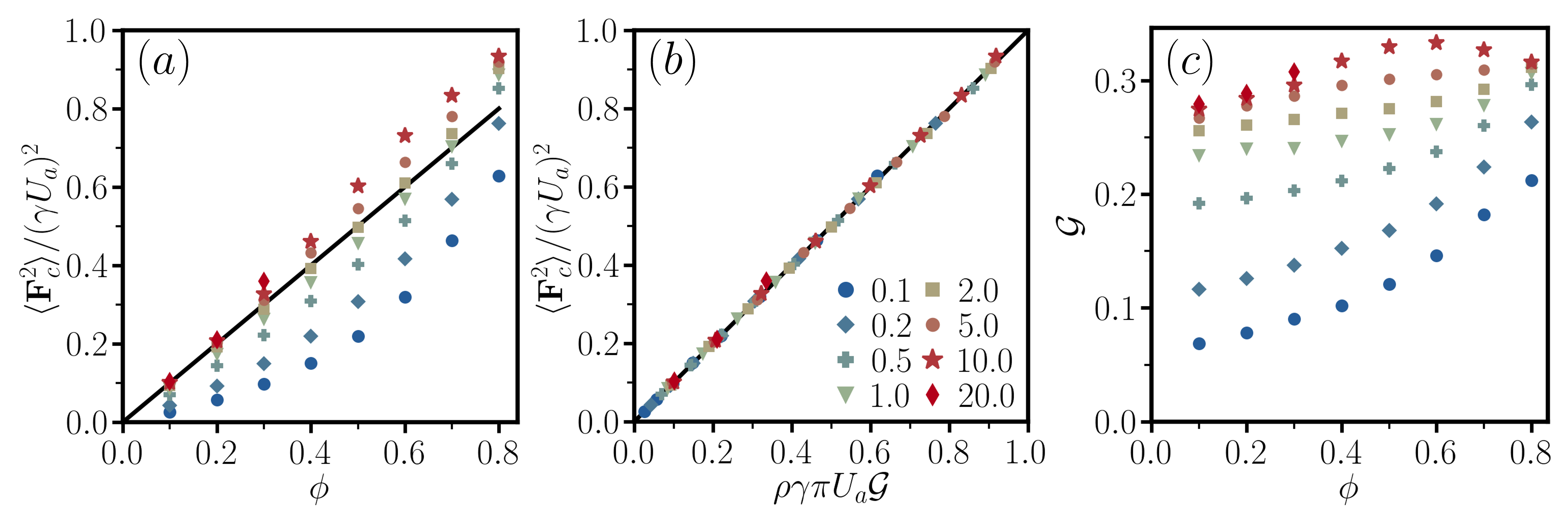}
\caption{
Structural origin of the collisional force variance.
(a) Collisional force variance $\langle \mathbf F_c^2\rangle$ as a function of area fraction $\phi$ for various P\'eclet numbers.
For $\mathrm{Pe}>1$, $\langle \mathbf F_c^2\rangle$ grows approximately linearly with \(\phi\); the solid black line corresponds to $\langle \mathbf F_c^2\rangle/(\gamma U_a)^2=\phi$.
(b) Parity plot comparing $\langle \mathbf F_c^2\rangle$ measured directly from simulations to the theoretical prediction based on the anisotropic pair correlation function [Eq.~(\ref{eq:16})].
The collapse onto the parity line demonstrates that the collisional force variance is quantitatively determined by local pair structure with no adjustable parameters.
(c) Force-weighted dipolar anisotropy $\mathcal{G}$ as a function of $\phi$ for various P\'eclet numbers.
For $\mathrm{Pe}>1$, $\mathcal{G}$ depends only weakly on \(\phi\), indicating that increasing $\langle \mathbf F_c^2\rangle$ is primarily driven by the number of near-contact neighbors rather than changes in angular structure.
}
\label{fig:4}
\end{figure*}

\section{Discussion}

The dependence of the self-diffusivity on \(\mathrm{Pe}\) and \(\phi\) can be codified using Eq.~(\ref{eq:13}), which separates the suppression of transport into two distinct contributions: a reduction of particle motion encoded by the collisional force variance \(\langle \mathbf F_c^2\rangle\), and a contribution arising from collisional correlations, quantified by the timescale \(\tau_D\).

Figure~\ref{fig:4}(a) shows the collisional force variance as a function of \(\mathrm{Pe}\) and \(\phi\).
For all activities, \(\langle \mathbf F_c^2\rangle\) increases monotonically with \(\phi\), reflecting the increasing number of collisions as the system becomes more crowded. 
For $\mathrm{Pe}<1$, particle orientations decorrelate rapidly, which limits the degree to which collisional forces are sustained, resulting in a weaker increase of \(\langle \mathbf F_c^2\rangle\) as a function of \(\phi\).
In contrast, for \(\mathrm{Pe}>1\), \(\langle \mathbf F_c^2\rangle\) exhibits an approximately linear dependence over a broad range of area fractions (i.e., \(\langle \mathbf F_c^2\rangle/(\gamma U_a)^2 \simeq \phi\)).

Figure~\ref{fig:4}(b) confirms the structural origin of the collisional force variance by comparing \(\langle \mathbf F_c^2\rangle\) measured from simulation to the prediction based on the anisotropic pair correlation function [Eq.~(\ref{eq:16})].
In Fig.~\ref{fig:4}(b), the data collapse onto the parity line over the full range of \(\mathrm{Pe}\) and \(\phi\) considered, confirming that the growth of \(\langle \mathbf F_c^2\rangle\) is quantitatively captured by the dipolar anisotropy of the local pair structure with no adjustable parameters.
In the Supplementary Material~\footnote{See Supplemental Material at [URL] for details.}, we provide representative examples of the pair correlation function \(g(r, \Phi)\).

To further elucidate the structural contribution, Fig.~\ref{fig:4}(c) isolates the force-weighted dipolar anisotropy of Eq.~(\ref{eq:16}) by defining
\begin{equation}
\label{eq:18}
\mathcal{G}
\equiv
\frac{\langle \mathbf F_c^2\rangle}{\rho \gamma U_a\,\pi}
=
\int_0^{\infty}
V'(r)\,
g_1(r)\,
r\,dr
 \, .
\end{equation}
For \(\mathrm{Pe}>1\), the quantity \(\mathcal{G}\) exhibits only a weak dependence on \(\phi\), indicating that the anisotropic structure established during collisions is largely insensitive to \(\phi\) in this regime.
Within this picture, the near-linear scaling \(\langle \mathbf F_c^2\rangle/(\gamma U_a)^2 \simeq \phi\) arises from a simple geometric effect: increasing \(\phi \) primarily increases the number of near-contact neighbors, while the angular distribution of collision partners remains largely unchanged.

While \(\langle \mathbf F_c^2\rangle\) controls the instantaneous suppression of particle motion, the self-diffusivity also depends on the temporal persistence of these forces, which we quantify through the net collisional memory time \(\tau_D\).
In the Supplementary Material~\cite{Note1}, we provide representative examples of the normalized force-force correlations \(C_{ca}(t)\) and \(C_{cc}(t)\), which are integrated to obtain the force-force correlation timescales \(\tau_{ca}\), \(\tau_{cc}\), and \(\tau_D\) [see Eq.~(\ref{eq:12})]. 

\begin{figure*}[t!]
\centering
\includegraphics[width=0.9\linewidth]{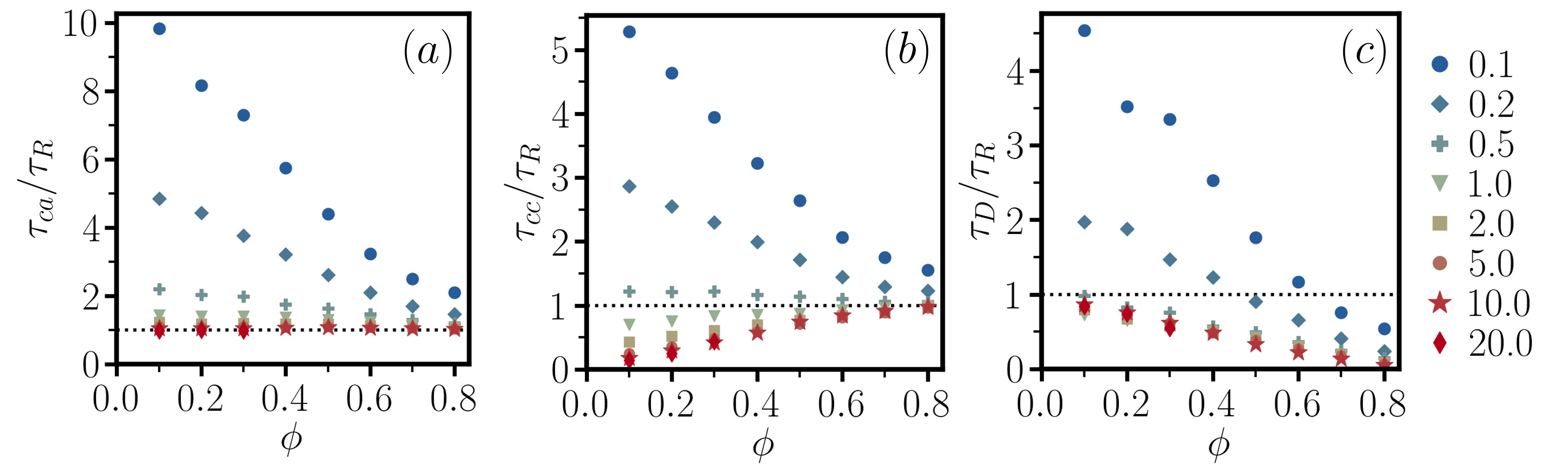}
\caption{Collisional correlation timescales modulating the self-diffusivity.
(a) Collisional force-orientation correlation time \(\tau_{ca}\). 
(b) Collisional force autocorrelation time \(\tau_{cc}\). 
(c) Net collisional memory timescale $\tau_D = \tau_{ca} - \tau_{cc}$, which quantifies the contribution of collisional memory to long-time transport.}
\label{fig:5}
\end{figure*}

In Fig.~\ref{fig:5}(a), we examine the collisional force-orientation correlation time \(\tau_{ca}\), which quantifies the persistence of collisional forces projected along the propulsion direction.
Across the representative \(\mathrm{Pe}-\phi\) state space, we find \(\tau_{ca}\ge\tau_R\), indicating that collisional forces remain correlated with the initial particle orientation for durations comparable to or exceeding the intrinsic reorientation time \(\tau_R\).

Notably, a clear distinction between low- and high-\(\mathrm{Pe}\) regimes emerges, corresponding to whether the run length is shorter or longer than the particle diameter.
For \(\mathrm{Pe}<1\), \(\tau_{ca}\) significantly exceeds \(\tau_R\), particularly at small \(\phi\).
This observation indicates that the decay of force–orientation correlations is governed by the lifetimes of local force-bearing configurations, rather than solely by rotational Brownian motion.
As \(\phi\) increases, configurational rearrangements accelerate, leading to a monotonic decrease of \(\tau_{ca}\) with \(\phi\).

In contrast, for \(\mathrm{Pe}>1\), we find that \(\tau_{ca}\approx\tau_R\) and depends only weakly on \(\phi\).
Once the run length exceeds the particle diameter, collisions are resolved at nearly fixed orientation, and the decay of collisional force-orientation correlations is controlled primarily by rotational diffusion.
Lastly, in the limit \(\phi \rightarrow \phi_c\), \(\tau_{ca} \rightarrow \tau_R\) for all \(\mathrm{Pe}\).
We provide a derivation in the Supplementary Material~\cite{Note1} that makes this limiting behavior explicit.

We next turn to the collisional force autocorrelation time \(\tau_{cc}\), shown in Fig.~\ref{fig:5}(b), which characterizes the persistence of the collisional force itself.
Unlike \(\tau_{ca}\), which couples force memory to orientation, \(\tau_{cc}\) reflects the timescale over which local force-bearing configurations decorrelate.
Accordingly, \(\tau_{cc}\) should be interpreted as a configurational rearrangement time associated with the renewal of the local contact environment, rather than as the duration of an individual binary collision as in dilute kinetic theories or simple crowding effects.

Similar to \(\tau_{ca}\), we find the behavior of \(\tau_{cc}\) exhibits two distinct regimes separated by \(\mathrm{Pe}\approx 1\).
For \(\mathrm{Pe}<1\), \(\tau_{cc}\) exceeds the intrinsic reorientation time \(\tau_R\), and decreases monotonically with increasing \(\phi\).
Intuitively, this suggests that increasing \(\phi\) enhances particle encounters and accelerates local rearrangements, leading to a reduction of \(\tau_{cc}\).
In contrast, for \(\mathrm{Pe}>1\), \(\tau_{cc}\) is upper bounded by \(\tau_R\) and increases monotonically with area fraction.
In this regime, force decorrelation is likely governed by the rate at which particles advect relative to their neighbors and exchange contacts.
As a consequence, \(\tau_{cc}/\tau_R\) nearly collapses for \(\mathrm{Pe} > 1\) and exhibits a simple \(\phi\) dependence.
In addition, in the limit \(\phi \rightarrow \phi_c\), \(\tau_{cc} \rightarrow \tau_R\) for all \(\mathrm{Pe}\), reflecting the fact that at large area fractions, frequent collisions rapidly renew local force-bearing configurations, leaving rotational diffusion as the dominant timescale.

Finally, Fig.~\ref{fig:5}(c) shows the net collisional memory timescale
\(\tau_D=\tau_{ca}-\tau_{cc}\),
which quantifies the total contribution of collisional memory to the self-diffusivity.
Across the representative \(\mathrm{Pe}-\phi\) state space, we find \(\tau_D\ge 0\), with equality approached as \(\phi \rightarrow \phi_c\).
Consistent with the trends discussed above, \(\tau_D\) is largest at low activity and small \(\phi\), where configurational dynamics and local environments are long-lived.
In this regime, collisional forces inherit memory from a slowly evolving structure.
In contrast, for \(\mathrm{Pe}>1\), particle motion rapidly renews local contact configurations, with \(\tau_D\) approaching \(\tau_R\) as \(\phi \rightarrow 0\) and tending to zero as \(\phi \rightarrow \phi_c\).

The nonnegativity of \(\tau_D\) reflects a separation between the timescales governing collisional force-orientation correlations and force autocorrelations.
As shown in Fig.~\ref{fig:5}, \(\tau_{ca} \geq \tau_{cc}\) for all \(\mathrm{Pe}\), indicating that collisional forces remain biased along the propulsion direction longer than they persist in magnitude alone.
Physically, this asymmetry arises from self-propulsion, which encodes a directional bias on successive collisions even as individual force-bearing contacts rearrange.

Importantly, the positivity of \(\tau_D\) implies that collisional memory cannot enhance self-diffusion in the purely repulsive active Brownian liquids considered here.
Instead, memory acts as an additional dissipative contribution, further reducing the self-diffusivity beyond that expected from equal-time speed renormalization alone.

\begin{figure*}[t!]
\centering
\includegraphics[width=0.9\linewidth]{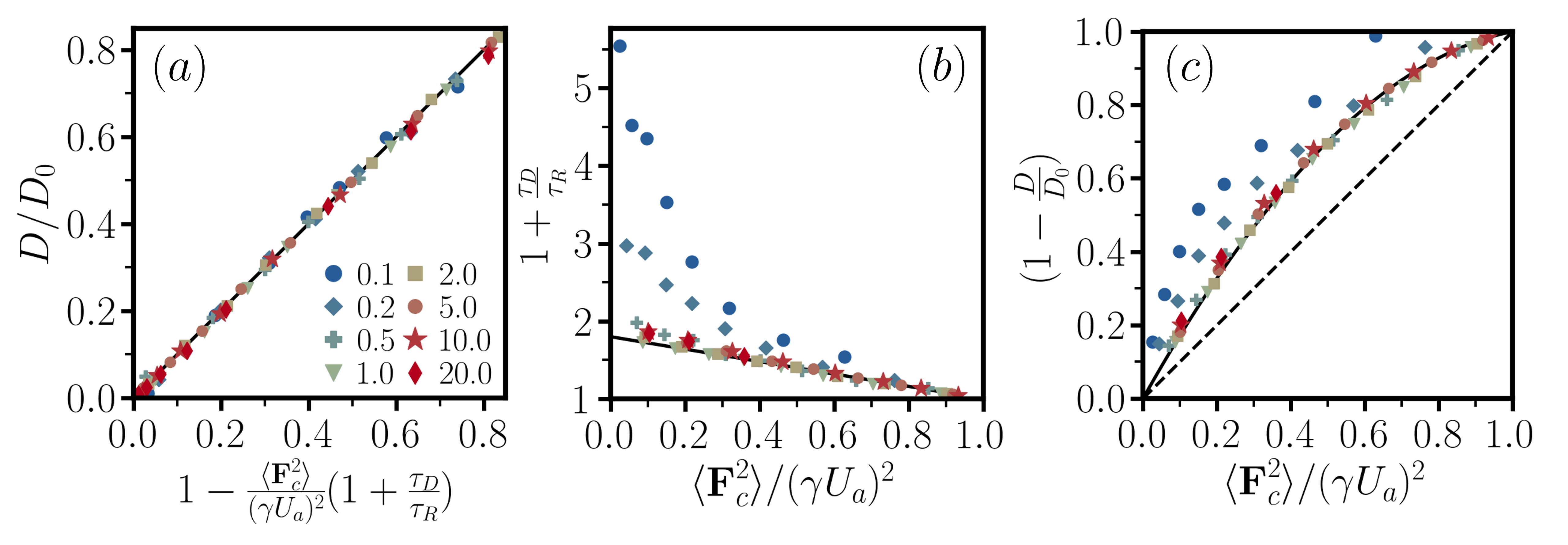}
\caption{
(a) Comparison of the measured self-diffusivity \(D/D_0\) with the theoretical prediction of Eq.~(\ref{eq:13}).
All data collapse onto the parity line, demonstrating that collisional force variance and collisional memory account for the observed transport behavior.
(b) Relative collisional memory factor \(1+\tau_D/\tau_R\) as a function of the normalized collisional force variance
\(\langle \mathbf F_c^2\rangle/(\gamma U_a)^2\) for various \(\mathrm{Pe}\).
For \(\mathrm{Pe}\gtrsim1\), the data collapse onto a nearly universal curve, given by
\(1+\tau_D/\tau_R = 1.8 - 0.8\,\langle \mathbf F_c^2\rangle/(\gamma U_a)^2\) (solid line).
(c) Reduction of the self-diffusivity, \(1-D/D_0\), versus the normalized collisional force variance.
The dashed line shows equal-time renormalization, while the solid line includes the memory correction inferred from (b).
Together, these results demonstrate that self-diffusion in interacting active Brownian liquids is quantitatively determined by the collisional force variance and a single memory timescale, with no adjustable parameters.
}
\label{fig:6}
\end{figure*}

Figure~\ref{fig:6} summarizes how collisional force variance and the net collisional memory combine to determine the self-diffusivity.
We begin with Figure~\ref{fig:6}(a), which directly tests the exact theoretical expression for the self-diffusivity [Eq.~(\ref{eq:13})].
Here, the measured self-diffusivity is compared to the prediction
\begin{equation*}
\frac{D}{D_0}
=
1
-
\frac{\langle \mathbf F_c^2\rangle}{(\gamma U_a)^2}
\left(
1+\frac{\tau_D}{\tau_R}
\right),
\end{equation*}
with both \(\langle \mathbf F_c^2\rangle\) and \(\tau_D\) obtained independently from simulation.
All data collapse onto the parity line with no adjustable parameters, confirming that the combined effects of the collisional force variance and collisional memory fully account for the observed transport behavior.

To better understand the dependence of the collisional force variance and the net collisional memory, we plot the relative collisional memory factor,
\(1+\tau_D/\tau_R\),
as a function of the normalized collisional force variance
\(\langle \mathbf F_c^2\rangle/(\gamma U_a)^2\) in Fig.~\ref{fig:6}(b).
For \(\mathrm{Pe}<1\), this quantity is large at small values of
\(\langle \mathbf F_c^2\rangle/(\gamma U_a)^2\)
and decreases as the force variance increases, reflecting the presence of long-lived configurational memory when particle motion is only weakly persistent.

For \(\mathrm{Pe}\gtrsim1\), the data collapse onto a nearly universal curve well described by the empirical linear form
\begin{equation}
\label{eq:19}
1+\frac{\tau_D}{\tau_R}
\;\approx\;
1.8
-
0.8\,\frac{\langle \mathbf F_c^2\rangle}{(\gamma U_a)^2},
\end{equation}
shown as the solid black line in Fig.~\ref{fig:6}(b).
This behavior of the memory factor \(1+\tau_D/\tau_R\) follows directly from the observed scaling of the two underlying correlation times.
The force-orientation correlation time exhibits a weak \(\phi\) dependence and remains close to the intrinsic reorientation time, with \(\tau_{ca}/\tau_R \sim \mathcal{O}(1)\), whereas the force autocorrelation time \(\tau_{cc}\) increases systematically with \(\phi\) and, equivalently, with the normalized collisional force variance \(\langle \mathbf F_c^2\rangle/(\gamma U_a)^2\).
Combining these trends via \(1+\tau_D/\tau_R = 1 + \tau_{ca}/\tau_R - \tau_{cc}/\tau_R\) explains both the collapse of the data and the approximately linear decrease of the memory factor with increasing \(\langle \mathbf F_c^2\rangle/(\gamma U_a)^2\).

Figure~\ref{fig:6}(c) compares the reduction of the self-diffusivity,
\(1-D/D_0\),
directly to the normalized collisional force variance.
Across the representative \(\mathrm{Pe}-\phi\) state space studied, we find
\begin{equation}
\label{eq:20}
1-\frac{D}{D_0}
\;\ge\;
\frac{\langle \mathbf F_c^2\rangle}{(\gamma U_a)^2},
\end{equation}
This inequality provides a complementary view of the parity test in Fig.~\ref{fig:6}(a), emphasizing that collisional memory necessarily contributes to the reduction of self-diffusivity.
For \(\mathrm{Pe}\gtrsim1\), these temporal corrections follow the same universal trend identified in Fig.~\ref{fig:6}(b), allowing the reduction in self-diffusivity to be approximated by
\begin{equation}
\label{eq:21}
1-\frac{D}{D_0}
\;\approx\;
\frac{\langle \mathbf F_c^2\rangle}{(\gamma U_a)^2}
\left[
1.8
-
0.8\,\frac{\langle \mathbf F_c^2\rangle}{(\gamma U_a)^2}
\right],
\end{equation}
shown as the solid black line in Fig.~\ref{fig:6}(c).
Using Eqs.~(\ref{eq:16}) and (\ref{eq:18}), this approximate expression can be written as
\begin{equation}
\label{eq:22}
1-\frac{D}{D_0}
\;\approx\;
\frac{\pi \rho \mathcal{G}}{\gamma U_a}
\left[
1.8
-
0.8\,\frac{\pi \rho \mathcal{G}}{\gamma U_a}
\right],
\end{equation}
where the deviation of the self-diffusivity is directly related to the number density \(\rho\) and the force-weighted dipolar anisotropy \(\mathcal{G}\), shown in Fig.~\ref{fig:4}(c).

Overall, Fig.~\ref{fig:6} demonstrates that the self-diffusivity of active Brownian liquids is fully controlled by two collisional quantities: the equal-time force variance and an additional timescale capturing collisional memory.
While the force variance sets the baseline reduction in transport, the memory factor \(1+\tau_D/\tau_R\) provides a systematic correction for temporal correlations.
For \(\mathrm{Pe} > 1\), these effects combine in a simple and nearly universal manner, yielding a compact description of self-diffusion in terms of measurable microscopic observables.

\section*{Conclusion}

We have developed a microscopic framework for self-diffusion in active Brownian liquids that organizes long-time transport in terms of collisional force statistics and their temporal persistence.
The central result of this work is the exact steady-state relation Eq.~(\ref{eq:13}), which provides a compact lens for understanding how interparticle interactions regulate transport.
Within this framework, deviations from ideal active diffusion arise from an instantaneous suppression of particle motion, encoded by the collisional force variance, and from a distinct contribution associated with collisional memory.

Applying this framework to purely repulsive ABPs, we find that equal-time force fluctuations provide the dominant reduction in self-diffusion.
Collisional memory introduces a correction that is always dissipative.
As a result, interparticle interactions cannot enhance self-diffusion in this class of systems.

A key physical insight emerging from this analysis is a separation between force generation and force renewal.
The collisional force variance reflects how near-contact structure generates anisotropic forces that oppose propulsion, whereas the memory contribution encodes how rapidly local force-bearing configurations are renewed by particle motion.
From this perspective, self-diffusion is governed not only by the strength of collisional forces but also by the rate at which force-bearing environments evolve.

Although our simulations focus on purely repulsive ABPs, the underlying framework is more general.
The decomposition of transport into equal-time force fluctuations and collisional memory provides a natural starting point for analyzing diffusion in more complex active systems, where interactions, environmental heterogeneity, or coupled orientational dynamics may alter the balance between force generation and renewal.

\appendix

\section{Simulation Details}

Simulations are performed using \texttt{HOOMD-blue} with a first-order overdamped integrator and a time step of $\delta t = 10^{-5}$ in reduced units.
Unless stated otherwise, we simulate systems of $\mathcal{N}=1000$ particles, running each simulation for $3\times 10^{9}$ time steps.
The first $5\times 10^{8}$ steps are discarded for equilibration, and data are sampled at least every $10^{2}$ steps thereafter.
The system is enclosed in a square box of area $\mathcal{A}$ with periodic boundary conditions.

Interparticle interactions are modeled using a generalized Lennard--Jones (Mie) potential with energy scale $\varepsilon$, diameter $\sigma$, repulsive exponent $n=50$, and attractive exponent $m=6$.
The potential is truncated and shifted at its minimum, yielding a purely repulsive interaction with an effective particle diameter $\sigma_p \approx 1.05\,\sigma$.
To ensure the hard-particle limit, we choose $\varepsilon/(\gamma \sigma U_a)=100$; further increases in $\varepsilon$ do not affect the results.
We have verified that all reported results are insensitive to system size and sampling frequency within the ranges explored.

%\bibliography{references}

\begin{thebibliography}{73}%
\makeatletter
\providecommand \@ifxundefined [1]{%
 \@ifx{#1\undefined}
}%
\providecommand \@ifnum [1]{%
 \ifnum #1\expandafter \@firstoftwo
 \else \expandafter \@secondoftwo
 \fi
}%
\providecommand \@ifx [1]{%
 \ifx #1\expandafter \@firstoftwo
 \else \expandafter \@secondoftwo
 \fi
}%
\providecommand \natexlab [1]{#1}%
\providecommand \enquote  [1]{``#1''}%
\providecommand \bibnamefont  [1]{#1}%
\providecommand \bibfnamefont [1]{#1}%
\providecommand \citenamefont [1]{#1}%
\providecommand \href@noop [0]{\@secondoftwo}%
\providecommand \href [0]{\begingroup \@sanitize@url \@href}%
\providecommand \@href[1]{\@@startlink{#1}\@@href}%
\providecommand \@@href[1]{\endgroup#1\@@endlink}%
\providecommand \@sanitize@url [0]{\catcode `\\12\catcode `\$12\catcode `\&12\catcode `\#12\catcode `\^12\catcode `\_12\catcode `\%12\relax}%
\providecommand \@@startlink[1]{}%
\providecommand \@@endlink[0]{}%
\providecommand \url  [0]{\begingroup\@sanitize@url \@url }%
\providecommand \@url [1]{\endgroup\@href {#1}{\urlprefix }}%
\providecommand \urlprefix  [0]{URL }%
\providecommand \Eprint [0]{\href }%
\providecommand \doibase [0]{https://doi.org/}%
\providecommand \selectlanguage [0]{\@gobble}%
\providecommand \bibinfo  [0]{\@secondoftwo}%
\providecommand \bibfield  [0]{\@secondoftwo}%
\providecommand \translation [1]{[#1]}%
\providecommand \BibitemOpen [0]{}%
\providecommand \bibitemStop [0]{}%
\providecommand \bibitemNoStop [0]{.\EOS\space}%
\providecommand \EOS [0]{\spacefactor3000\relax}%
\providecommand \BibitemShut  [1]{\csname bibitem#1\endcsname}%
\let\auto@bib@innerbib\@empty
%</preamble>
\bibitem [{\citenamefont {Dhont}(1996)}]{dhont1996}%
  \BibitemOpen
  \bibfield  {author} {\bibinfo {author} {\bibfnamefont {J.}~\bibnamefont {Dhont}},\ }\href@noop {} {\emph {\bibinfo {title} {An Introduction to Dynamics of Colloids}}},\ Studies in Interface Science\ (\bibinfo  {publisher} {Elsevier Science},\ \bibinfo {year} {1996})\BibitemShut {NoStop}%
\bibitem [{\citenamefont {Chapman}\ and\ \citenamefont {Cowling}(1990)}]{chapman1990mathematical}%
  \BibitemOpen
  \bibfield  {author} {\bibinfo {author} {\bibfnamefont {S.}~\bibnamefont {Chapman}}\ and\ \bibinfo {author} {\bibfnamefont {T.}~\bibnamefont {Cowling}},\ }\href@noop {} {\emph {\bibinfo {title} {The Mathematical Theory of Non-uniform Gases: An Account of the Kinetic Theory of Viscosity, Thermal Conduction and Diffusion in Gases}}},\ Cambridge Mathematical Library\ (\bibinfo  {publisher} {Cambridge University Press},\ \bibinfo {year} {1990})\BibitemShut {NoStop}%
\bibitem [{\citenamefont {Zahn}\ \emph {et~al.}(1997)\citenamefont {Zahn}, \citenamefont {Méndez-Alcaraz},\ and\ \citenamefont {Maret}}]{Zahn1997-ne}%
  \BibitemOpen
  \bibfield  {author} {\bibinfo {author} {\bibfnamefont {K.}~\bibnamefont {Zahn}}, \bibinfo {author} {\bibfnamefont {J.~M.}\ \bibnamefont {Méndez-Alcaraz}},\ and\ \bibinfo {author} {\bibfnamefont {G.}~\bibnamefont {Maret}},\ }\bibfield  {title} {\bibinfo {title} {Hydrodynamic interactions may enhance the self-diffusion of colloidal particles},\ }\href@noop {} {\bibfield  {journal} {\bibinfo  {journal} {Phys. Rev. Lett.}\ }\textbf {\bibinfo {volume} {79}},\ \bibinfo {pages} {175} (\bibinfo {year} {1997})}\BibitemShut {NoStop}%
\bibitem [{\citenamefont {Zwanzig}(2001)}]{zwanzig2001nonequilibrium}%
  \BibitemOpen
  \bibfield  {author} {\bibinfo {author} {\bibfnamefont {R.}~\bibnamefont {Zwanzig}},\ }\href@noop {} {\emph {\bibinfo {title} {Nonequilibrium Statistical Mechanics}}}\ (\bibinfo  {publisher} {Oxford University Press},\ \bibinfo {year} {2001})\BibitemShut {NoStop}%
\bibitem [{\citenamefont {Bian}\ \emph {et~al.}(2016)\citenamefont {Bian}, \citenamefont {Kim},\ and\ \citenamefont {Karniadakis}}]{Bian2016-xu}%
  \BibitemOpen
  \bibfield  {author} {\bibinfo {author} {\bibfnamefont {X.}~\bibnamefont {Bian}}, \bibinfo {author} {\bibfnamefont {C.}~\bibnamefont {Kim}},\ and\ \bibinfo {author} {\bibfnamefont {G.~E.}\ \bibnamefont {Karniadakis}},\ }\bibfield  {title} {\bibinfo {title} {111 years of brownian motion},\ }\href@noop {} {\bibfield  {journal} {\bibinfo  {journal} {Soft Matter}\ }\textbf {\bibinfo {volume} {12}},\ \bibinfo {pages} {6331} (\bibinfo {year} {2016})}\BibitemShut {NoStop}%
\bibitem [{\citenamefont {Balakrishnan}(2020)}]{balakrishnan2020elements}%
  \BibitemOpen
  \bibfield  {author} {\bibinfo {author} {\bibfnamefont {V.}~\bibnamefont {Balakrishnan}},\ }\href@noop {} {\emph {\bibinfo {title} {Elements of Nonequilibrium Statistical Mechanics}}}\ (\bibinfo  {publisher} {Springer International Publishing},\ \bibinfo {year} {2020})\BibitemShut {NoStop}%
\bibitem [{\citenamefont {Weeks}\ and\ \citenamefont {Weitz}(2002)}]{Weeks2002-lo}%
  \BibitemOpen
  \bibfield  {author} {\bibinfo {author} {\bibfnamefont {E.~R.}\ \bibnamefont {Weeks}}\ and\ \bibinfo {author} {\bibfnamefont {D.~A.}\ \bibnamefont {Weitz}},\ }\bibfield  {title} {\bibinfo {title} {Properties of cage rearrangements observed near the colloidal glass transition},\ }\href@noop {} {\bibfield  {journal} {\bibinfo  {journal} {Phys. Rev. Lett.}\ }\textbf {\bibinfo {volume} {89}},\ \bibinfo {pages} {095704} (\bibinfo {year} {2002})}\BibitemShut {NoStop}%
\bibitem [{\citenamefont {Berthier}\ and\ \citenamefont {Biroli}(2011)}]{Berthier2011-sc}%
  \BibitemOpen
  \bibfield  {author} {\bibinfo {author} {\bibfnamefont {L.}~\bibnamefont {Berthier}}\ and\ \bibinfo {author} {\bibfnamefont {G.}~\bibnamefont {Biroli}},\ }\bibfield  {title} {\bibinfo {title} {Theoretical perspective on the glass transition and amorphous materials},\ }\href@noop {} {\bibfield  {journal} {\bibinfo  {journal} {Rev. Mod. Phys.}\ }\textbf {\bibinfo {volume} {83}},\ \bibinfo {pages} {587} (\bibinfo {year} {2011})}\BibitemShut {NoStop}%
\bibitem [{\citenamefont {Hunter}\ and\ \citenamefont {Weeks}(2012)}]{Hunter2012-qf}%
  \BibitemOpen
  \bibfield  {author} {\bibinfo {author} {\bibfnamefont {G.~L.}\ \bibnamefont {Hunter}}\ and\ \bibinfo {author} {\bibfnamefont {E.~R.}\ \bibnamefont {Weeks}},\ }\bibfield  {title} {\bibinfo {title} {The physics of the colloidal glass transition},\ }\href@noop {} {\bibfield  {journal} {\bibinfo  {journal} {Rep. Prog. Phys.}\ }\textbf {\bibinfo {volume} {75}},\ \bibinfo {pages} {066501} (\bibinfo {year} {2012})}\BibitemShut {NoStop}%
\bibitem [{\citenamefont {Berthier}(2014)}]{Berthier2014-la}%
  \BibitemOpen
  \bibfield  {author} {\bibinfo {author} {\bibfnamefont {L.}~\bibnamefont {Berthier}},\ }\bibfield  {title} {\bibinfo {title} {Nonequilibrium glassy dynamics of self-propelled hard disks},\ }\href@noop {} {\bibfield  {journal} {\bibinfo  {journal} {Phys. Rev. Lett.}\ }\textbf {\bibinfo {volume} {112}},\ \bibinfo {pages} {220602} (\bibinfo {year} {2014})}\BibitemShut {NoStop}%
\bibitem [{\citenamefont {Romanczuk}\ \emph {et~al.}(2012)\citenamefont {Romanczuk}, \citenamefont {Bär}, \citenamefont {Ebeling}, \citenamefont {Lindner},\ and\ \citenamefont {Schimansky-Geier}}]{Romanczuk2012-dw}%
  \BibitemOpen
  \bibfield  {author} {\bibinfo {author} {\bibfnamefont {P.}~\bibnamefont {Romanczuk}}, \bibinfo {author} {\bibfnamefont {M.}~\bibnamefont {Bär}}, \bibinfo {author} {\bibfnamefont {W.}~\bibnamefont {Ebeling}}, \bibinfo {author} {\bibfnamefont {B.}~\bibnamefont {Lindner}},\ and\ \bibinfo {author} {\bibfnamefont {L.}~\bibnamefont {Schimansky-Geier}},\ }\bibfield  {title} {\bibinfo {title} {Active brownian particles},\ }\href@noop {} {\bibfield  {journal} {\bibinfo  {journal} {Eur. Phys. J. Spec. Top.}\ }\textbf {\bibinfo {volume} {202}},\ \bibinfo {pages} {1} (\bibinfo {year} {2012})}\BibitemShut {NoStop}%
\bibitem [{\citenamefont {Nardini}\ \emph {et~al.}(2017)\citenamefont {Nardini}, \citenamefont {Fodor}, \citenamefont {Tjhung}, \citenamefont {van Wijland}, \citenamefont {Tailleur},\ and\ \citenamefont {Cates}}]{Nardini2017-ni}%
  \BibitemOpen
  \bibfield  {author} {\bibinfo {author} {\bibfnamefont {C.}~\bibnamefont {Nardini}}, \bibinfo {author} {\bibfnamefont {E.}~\bibnamefont {Fodor}}, \bibinfo {author} {\bibfnamefont {E.}~\bibnamefont {Tjhung}}, \bibinfo {author} {\bibfnamefont {F.}~\bibnamefont {van Wijland}}, \bibinfo {author} {\bibfnamefont {J.}~\bibnamefont {Tailleur}},\ and\ \bibinfo {author} {\bibfnamefont {M.~E.}\ \bibnamefont {Cates}},\ }\bibfield  {title} {\bibinfo {title} {Entropy production in field theories without time-reversal symmetry: Quantifying the non-equilibrium character of active matter},\ }\href@noop {} {\bibfield  {journal} {\bibinfo  {journal} {Phys. Rev. X}\ }\textbf {\bibinfo {volume} {7}},\ \bibinfo {pages} {021007} (\bibinfo {year} {2017})}\BibitemShut {NoStop}%
\bibitem [{\citenamefont {Fodor}\ \emph {et~al.}(2016)\citenamefont {Fodor}, \citenamefont {Nardini}, \citenamefont {Cates}, \citenamefont {Tailleur}, \citenamefont {Visco},\ and\ \citenamefont {van Wijland}}]{Fodor2016-nb}%
  \BibitemOpen
  \bibfield  {author} {\bibinfo {author} {\bibfnamefont {E.}~\bibnamefont {Fodor}}, \bibinfo {author} {\bibfnamefont {C.}~\bibnamefont {Nardini}}, \bibinfo {author} {\bibfnamefont {M.~E.}\ \bibnamefont {Cates}}, \bibinfo {author} {\bibfnamefont {J.}~\bibnamefont {Tailleur}}, \bibinfo {author} {\bibfnamefont {P.}~\bibnamefont {Visco}},\ and\ \bibinfo {author} {\bibfnamefont {F.}~\bibnamefont {van Wijland}},\ }\bibfield  {title} {\bibinfo {title} {How far from equilibrium is active matter?},\ }\href@noop {} {\bibfield  {journal} {\bibinfo  {journal} {Phys. Rev. Lett.}\ }\textbf {\bibinfo {volume} {117}},\ \bibinfo {pages} {038103} (\bibinfo {year} {2016})}\BibitemShut {NoStop}%
\bibitem [{\citenamefont {Paxton}\ \emph {et~al.}(2004)\citenamefont {Paxton}, \citenamefont {Kistler}, \citenamefont {Olmeda}, \citenamefont {Sen}, \citenamefont {St~Angelo}, \citenamefont {Cao}, \citenamefont {Mallouk}, \citenamefont {Lammert},\ and\ \citenamefont {Crespi}}]{Paxton2004-vw}%
  \BibitemOpen
  \bibfield  {author} {\bibinfo {author} {\bibfnamefont {W.~F.}\ \bibnamefont {Paxton}}, \bibinfo {author} {\bibfnamefont {K.~C.}\ \bibnamefont {Kistler}}, \bibinfo {author} {\bibfnamefont {C.~C.}\ \bibnamefont {Olmeda}}, \bibinfo {author} {\bibfnamefont {A.}~\bibnamefont {Sen}}, \bibinfo {author} {\bibfnamefont {S.~K.}\ \bibnamefont {St~Angelo}}, \bibinfo {author} {\bibfnamefont {Y.}~\bibnamefont {Cao}}, \bibinfo {author} {\bibfnamefont {T.~E.}\ \bibnamefont {Mallouk}}, \bibinfo {author} {\bibfnamefont {P.~E.}\ \bibnamefont {Lammert}},\ and\ \bibinfo {author} {\bibfnamefont {V.~H.}\ \bibnamefont {Crespi}},\ }\bibfield  {title} {\bibinfo {title} {Catalytic nanomotors: autonomous movement of striped nanorods},\ }\href@noop {} {\bibfield  {journal} {\bibinfo  {journal} {J. Am. Chem. Soc.}\ }\textbf {\bibinfo {volume} {126}},\ \bibinfo {pages} {13424} (\bibinfo {year} {2004})}\BibitemShut {NoStop}%
\bibitem [{\citenamefont {Golestanian}\ \emph {et~al.}(2005)\citenamefont {Golestanian}, \citenamefont {Liverpool},\ and\ \citenamefont {Ajdari}}]{Golestanian2005-pf}%
  \BibitemOpen
  \bibfield  {author} {\bibinfo {author} {\bibfnamefont {R.}~\bibnamefont {Golestanian}}, \bibinfo {author} {\bibfnamefont {T.~B.}\ \bibnamefont {Liverpool}},\ and\ \bibinfo {author} {\bibfnamefont {A.}~\bibnamefont {Ajdari}},\ }\bibfield  {title} {\bibinfo {title} {Propulsion of a molecular machine by asymmetric distribution of reaction products},\ }\href@noop {} {\bibfield  {journal} {\bibinfo  {journal} {Phys. Rev. Lett.}\ }\textbf {\bibinfo {volume} {94}},\ \bibinfo {pages} {220801} (\bibinfo {year} {2005})}\BibitemShut {NoStop}%
\bibitem [{\citenamefont {Wang}\ \emph {et~al.}(2006)\citenamefont {Wang}, \citenamefont {Hernandez}, \citenamefont {Bartlett}, \citenamefont {Bingham}, \citenamefont {Kline}, \citenamefont {Sen},\ and\ \citenamefont {Mallouk}}]{Wang2006-je}%
  \BibitemOpen
  \bibfield  {author} {\bibinfo {author} {\bibfnamefont {Y.}~\bibnamefont {Wang}}, \bibinfo {author} {\bibfnamefont {R.~M.}\ \bibnamefont {Hernandez}}, \bibinfo {author} {\bibfnamefont {D.~J.}\ \bibnamefont {Bartlett}, \bibfnamefont {Jr}}, \bibinfo {author} {\bibfnamefont {J.~M.}\ \bibnamefont {Bingham}}, \bibinfo {author} {\bibfnamefont {T.~R.}\ \bibnamefont {Kline}}, \bibinfo {author} {\bibfnamefont {A.}~\bibnamefont {Sen}},\ and\ \bibinfo {author} {\bibfnamefont {T.~E.}\ \bibnamefont {Mallouk}},\ }\bibfield  {title} {\bibinfo {title} {Bipolar electrochemical mechanism for the propulsion of catalytic nanomotors in hydrogen peroxide solutions},\ }\href@noop {} {\bibfield  {journal} {\bibinfo  {journal} {Langmuir}\ }\textbf {\bibinfo {volume} {22}},\ \bibinfo {pages} {10451} (\bibinfo {year} {2006})}\BibitemShut {NoStop}%
\bibitem [{\citenamefont {Ghosh}\ and\ \citenamefont {Fischer}(2009)}]{Ghosh2009-ok}%
  \BibitemOpen
  \bibfield  {author} {\bibinfo {author} {\bibfnamefont {A.}~\bibnamefont {Ghosh}}\ and\ \bibinfo {author} {\bibfnamefont {P.}~\bibnamefont {Fischer}},\ }\bibfield  {title} {\bibinfo {title} {Controlled propulsion of artificial magnetic nanostructured propellers},\ }\href@noop {} {\bibfield  {journal} {\bibinfo  {journal} {Nano Lett.}\ }\textbf {\bibinfo {volume} {9}},\ \bibinfo {pages} {2243} (\bibinfo {year} {2009})}\BibitemShut {NoStop}%
\bibitem [{\citenamefont {Ebbens}\ and\ \citenamefont {Howse}(2010)}]{Ebbens2010-nb}%
  \BibitemOpen
  \bibfield  {author} {\bibinfo {author} {\bibfnamefont {S.~J.}\ \bibnamefont {Ebbens}}\ and\ \bibinfo {author} {\bibfnamefont {J.~R.}\ \bibnamefont {Howse}},\ }\bibfield  {title} {\bibinfo {title} {In pursuit of propulsion at the nanoscale},\ }\href@noop {} {\bibfield  {journal} {\bibinfo  {journal} {Soft Matter}\ }\textbf {\bibinfo {volume} {6}},\ \bibinfo {pages} {726} (\bibinfo {year} {2010})}\BibitemShut {NoStop}%
\bibitem [{\citenamefont {Gao}\ \emph {et~al.}(2012)\citenamefont {Gao}, \citenamefont {Uygun},\ and\ \citenamefont {Wang}}]{Gao2012-gc}%
  \BibitemOpen
  \bibfield  {author} {\bibinfo {author} {\bibfnamefont {W.}~\bibnamefont {Gao}}, \bibinfo {author} {\bibfnamefont {A.}~\bibnamefont {Uygun}},\ and\ \bibinfo {author} {\bibfnamefont {J.}~\bibnamefont {Wang}},\ }\bibfield  {title} {\bibinfo {title} {Hydrogen-bubble-propelled zinc-based microrockets in strongly acidic media},\ }\href@noop {} {\bibfield  {journal} {\bibinfo  {journal} {J. Am. Chem. Soc.}\ }\textbf {\bibinfo {volume} {134}},\ \bibinfo {pages} {897} (\bibinfo {year} {2012})}\BibitemShut {NoStop}%
\bibitem [{\citenamefont {Izri}\ \emph {et~al.}(2014)\citenamefont {Izri}, \citenamefont {van~der Linden}, \citenamefont {Michelin},\ and\ \citenamefont {Dauchot}}]{Izri2014-dh}%
  \BibitemOpen
  \bibfield  {author} {\bibinfo {author} {\bibfnamefont {Z.}~\bibnamefont {Izri}}, \bibinfo {author} {\bibfnamefont {M.~N.}\ \bibnamefont {van~der Linden}}, \bibinfo {author} {\bibfnamefont {S.}~\bibnamefont {Michelin}},\ and\ \bibinfo {author} {\bibfnamefont {O.}~\bibnamefont {Dauchot}},\ }\bibfield  {title} {\bibinfo {title} {Self-propulsion of pure water droplets by spontaneous marangoni-stress-driven motion},\ }\href@noop {} {\bibfield  {journal} {\bibinfo  {journal} {Phys. Rev. Lett.}\ }\textbf {\bibinfo {volume} {113}},\ \bibinfo {pages} {248302} (\bibinfo {year} {2014})}\BibitemShut {NoStop}%
\bibitem [{\citenamefont {Moran}\ and\ \citenamefont {Posner}(2017)}]{Moran2017-yk}%
  \BibitemOpen
  \bibfield  {author} {\bibinfo {author} {\bibfnamefont {J.~L.}\ \bibnamefont {Moran}}\ and\ \bibinfo {author} {\bibfnamefont {J.~D.}\ \bibnamefont {Posner}},\ }\bibfield  {title} {\bibinfo {title} {Phoretic self-propulsion},\ }\href@noop {} {\bibfield  {journal} {\bibinfo  {journal} {Annu. Rev. Fluid Mech.}\ }\textbf {\bibinfo {volume} {49}},\ \bibinfo {pages} {511} (\bibinfo {year} {2017})}\BibitemShut {NoStop}%
\bibitem [{\citenamefont {Romanczuk}\ and\ \citenamefont {Schimansky-Geier}(2011)}]{Romanczuk2011-ym}%
  \BibitemOpen
  \bibfield  {author} {\bibinfo {author} {\bibfnamefont {P.}~\bibnamefont {Romanczuk}}\ and\ \bibinfo {author} {\bibfnamefont {L.}~\bibnamefont {Schimansky-Geier}},\ }\bibfield  {title} {\bibinfo {title} {Brownian motion with active fluctuations},\ }\href@noop {} {\bibfield  {journal} {\bibinfo  {journal} {Phys. Rev. Lett.}\ }\textbf {\bibinfo {volume} {106}},\ \bibinfo {pages} {230601} (\bibinfo {year} {2011})}\BibitemShut {NoStop}%
\bibitem [{\citenamefont {Schweitzer}(2003)}]{schweitzer2003brownian}%
  \BibitemOpen
  \bibfield  {author} {\bibinfo {author} {\bibfnamefont {F.}~\bibnamefont {Schweitzer}},\ }\href@noop {} {\emph {\bibinfo {title} {Brownian Agents and Active Particles: Collective Dynamics in the Natural and Social Sciences}}},\ Physics and astronomy online library\ (\bibinfo  {publisher} {Springer Berlin Heidelberg},\ \bibinfo {year} {2003})\BibitemShut {NoStop}%
\bibitem [{\citenamefont {Nelson}\ \emph {et~al.}(2010)\citenamefont {Nelson}, \citenamefont {Kaliakatsos},\ and\ \citenamefont {Abbott}}]{Nelson2010-jh}%
  \BibitemOpen
  \bibfield  {author} {\bibinfo {author} {\bibfnamefont {B.~J.}\ \bibnamefont {Nelson}}, \bibinfo {author} {\bibfnamefont {I.~K.}\ \bibnamefont {Kaliakatsos}},\ and\ \bibinfo {author} {\bibfnamefont {J.~J.}\ \bibnamefont {Abbott}},\ }\bibfield  {title} {\bibinfo {title} {Microrobots for minimally invasive medicine},\ }\href@noop {} {\bibfield  {journal} {\bibinfo  {journal} {Annu. Rev. Biomed. Eng.}\ }\textbf {\bibinfo {volume} {12}},\ \bibinfo {pages} {55} (\bibinfo {year} {2010})}\BibitemShut {NoStop}%
\bibitem [{\citenamefont {Sanchez}\ \emph {et~al.}(2011)\citenamefont {Sanchez}, \citenamefont {Solovev}, \citenamefont {Schulze},\ and\ \citenamefont {Schmidt}}]{Sanchez2011-ej}%
  \BibitemOpen
  \bibfield  {author} {\bibinfo {author} {\bibfnamefont {S.}~\bibnamefont {Sanchez}}, \bibinfo {author} {\bibfnamefont {A.~A.}\ \bibnamefont {Solovev}}, \bibinfo {author} {\bibfnamefont {S.}~\bibnamefont {Schulze}},\ and\ \bibinfo {author} {\bibfnamefont {O.~G.}\ \bibnamefont {Schmidt}},\ }\bibfield  {title} {\bibinfo {title} {Controlled manipulation of multiple cells using catalytic microbots},\ }\href@noop {} {\bibfield  {journal} {\bibinfo  {journal} {Chem. Commun. (Camb.)}\ }\textbf {\bibinfo {volume} {47}},\ \bibinfo {pages} {698} (\bibinfo {year} {2011})}\BibitemShut {NoStop}%
\bibitem [{\citenamefont {Wang}\ and\ \citenamefont {Gao}(2012)}]{Wang2012-kd}%
  \BibitemOpen
  \bibfield  {author} {\bibinfo {author} {\bibfnamefont {J.}~\bibnamefont {Wang}}\ and\ \bibinfo {author} {\bibfnamefont {W.}~\bibnamefont {Gao}},\ }\bibfield  {title} {\bibinfo {title} {Nano/microscale motors: biomedical opportunities and challenges},\ }\href@noop {} {\bibfield  {journal} {\bibinfo  {journal} {ACS Nano}\ }\textbf {\bibinfo {volume} {6}},\ \bibinfo {pages} {5745} (\bibinfo {year} {2012})}\BibitemShut {NoStop}%
\bibitem [{\citenamefont {Soler}\ \emph {et~al.}(2013)\citenamefont {Soler}, \citenamefont {Magdanz}, \citenamefont {Fomin}, \citenamefont {Sanchez},\ and\ \citenamefont {Schmidt}}]{Soler2013-vk}%
  \BibitemOpen
  \bibfield  {author} {\bibinfo {author} {\bibfnamefont {L.}~\bibnamefont {Soler}}, \bibinfo {author} {\bibfnamefont {V.}~\bibnamefont {Magdanz}}, \bibinfo {author} {\bibfnamefont {V.~M.}\ \bibnamefont {Fomin}}, \bibinfo {author} {\bibfnamefont {S.}~\bibnamefont {Sanchez}},\ and\ \bibinfo {author} {\bibfnamefont {O.~G.}\ \bibnamefont {Schmidt}},\ }\bibfield  {title} {\bibinfo {title} {Self-propelled micromotors for cleaning polluted water},\ }\href@noop {} {\bibfield  {journal} {\bibinfo  {journal} {ACS Nano}\ }\textbf {\bibinfo {volume} {7}},\ \bibinfo {pages} {9611} (\bibinfo {year} {2013})}\BibitemShut {NoStop}%
\bibitem [{\citenamefont {Patra}\ \emph {et~al.}(2013)\citenamefont {Patra}, \citenamefont {Sengupta}, \citenamefont {Duan}, \citenamefont {Zhang}, \citenamefont {Pavlick},\ and\ \citenamefont {Sen}}]{Patra2013-by}%
  \BibitemOpen
  \bibfield  {author} {\bibinfo {author} {\bibfnamefont {D.}~\bibnamefont {Patra}}, \bibinfo {author} {\bibfnamefont {S.}~\bibnamefont {Sengupta}}, \bibinfo {author} {\bibfnamefont {W.}~\bibnamefont {Duan}}, \bibinfo {author} {\bibfnamefont {H.}~\bibnamefont {Zhang}}, \bibinfo {author} {\bibfnamefont {R.}~\bibnamefont {Pavlick}},\ and\ \bibinfo {author} {\bibfnamefont {A.}~\bibnamefont {Sen}},\ }\bibfield  {title} {\bibinfo {title} {Intelligent, self-powered, drug delivery systems},\ }\href@noop {} {\bibfield  {journal} {\bibinfo  {journal} {Nanoscale}\ }\textbf {\bibinfo {volume} {5}},\ \bibinfo {pages} {1273} (\bibinfo {year} {2013})}\BibitemShut {NoStop}%
\bibitem [{\citenamefont {Garcia-Gradilla}\ \emph {et~al.}(2013)\citenamefont {Garcia-Gradilla}, \citenamefont {Orozco}, \citenamefont {Sattayasamitsathit}, \citenamefont {Soto}, \citenamefont {Kuralay}, \citenamefont {Pourazary}, \citenamefont {Katzenberg}, \citenamefont {Gao}, \citenamefont {Shen},\ and\ \citenamefont {Wang}}]{Garcia-Gradilla2013-zk}%
  \BibitemOpen
  \bibfield  {author} {\bibinfo {author} {\bibfnamefont {V.}~\bibnamefont {Garcia-Gradilla}}, \bibinfo {author} {\bibfnamefont {J.}~\bibnamefont {Orozco}}, \bibinfo {author} {\bibfnamefont {S.}~\bibnamefont {Sattayasamitsathit}}, \bibinfo {author} {\bibfnamefont {F.}~\bibnamefont {Soto}}, \bibinfo {author} {\bibfnamefont {F.}~\bibnamefont {Kuralay}}, \bibinfo {author} {\bibfnamefont {A.}~\bibnamefont {Pourazary}}, \bibinfo {author} {\bibfnamefont {A.}~\bibnamefont {Katzenberg}}, \bibinfo {author} {\bibfnamefont {W.}~\bibnamefont {Gao}}, \bibinfo {author} {\bibfnamefont {Y.}~\bibnamefont {Shen}},\ and\ \bibinfo {author} {\bibfnamefont {J.}~\bibnamefont {Wang}},\ }\bibfield  {title} {\bibinfo {title} {Functionalized ultrasound-propelled magnetically guided nanomotors: toward practical biomedical applications},\ }\href@noop {} {\bibfield  {journal} {\bibinfo  {journal} {ACS Nano}\ }\textbf {\bibinfo {volume} {7}},\ \bibinfo {pages} {9232} (\bibinfo {year} {2013})}\BibitemShut {NoStop}%
\bibitem [{\citenamefont {Guix}\ \emph {et~al.}(2014)\citenamefont {Guix}, \citenamefont {Mayorga-Martinez},\ and\ \citenamefont {Merkoçi}}]{Guix2014-jz}%
  \BibitemOpen
  \bibfield  {author} {\bibinfo {author} {\bibfnamefont {M.}~\bibnamefont {Guix}}, \bibinfo {author} {\bibfnamefont {C.~C.}\ \bibnamefont {Mayorga-Martinez}},\ and\ \bibinfo {author} {\bibfnamefont {A.}~\bibnamefont {Merkoçi}},\ }\bibfield  {title} {\bibinfo {title} {Nano/micromotors in (bio)chemical science applications},\ }\href@noop {} {\bibfield  {journal} {\bibinfo  {journal} {Chem. Rev.}\ }\textbf {\bibinfo {volume} {114}},\ \bibinfo {pages} {6285} (\bibinfo {year} {2014})}\BibitemShut {NoStop}%
\bibitem [{\citenamefont {Gao}\ and\ \citenamefont {Wang}(2014)}]{Gao2014-bq}%
  \BibitemOpen
  \bibfield  {author} {\bibinfo {author} {\bibfnamefont {W.}~\bibnamefont {Gao}}\ and\ \bibinfo {author} {\bibfnamefont {J.}~\bibnamefont {Wang}},\ }\bibfield  {title} {\bibinfo {title} {The environmental impact of micro/nanomachines: a review},\ }\href@noop {} {\bibfield  {journal} {\bibinfo  {journal} {ACS Nano}\ }\textbf {\bibinfo {volume} {8}},\ \bibinfo {pages} {3170} (\bibinfo {year} {2014})}\BibitemShut {NoStop}%
\bibitem [{\citenamefont {Li}\ \emph {et~al.}(2016)\citenamefont {Li}, \citenamefont {Rozen},\ and\ \citenamefont {Wang}}]{Li2016-yv}%
  \BibitemOpen
  \bibfield  {author} {\bibinfo {author} {\bibfnamefont {J.}~\bibnamefont {Li}}, \bibinfo {author} {\bibfnamefont {I.}~\bibnamefont {Rozen}},\ and\ \bibinfo {author} {\bibfnamefont {J.}~\bibnamefont {Wang}},\ }\bibfield  {title} {\bibinfo {title} {Rocket science at the nanoscale},\ }\href@noop {} {\bibfield  {journal} {\bibinfo  {journal} {ACS Nano}\ }\textbf {\bibinfo {volume} {10}},\ \bibinfo {pages} {5619} (\bibinfo {year} {2016})}\BibitemShut {NoStop}%
\bibitem [{\citenamefont {Needleman}\ and\ \citenamefont {Dogic}(2017)}]{Needleman2017-lh}%
  \BibitemOpen
  \bibfield  {author} {\bibinfo {author} {\bibfnamefont {D.}~\bibnamefont {Needleman}}\ and\ \bibinfo {author} {\bibfnamefont {Z.}~\bibnamefont {Dogic}},\ }\bibfield  {title} {\bibinfo {title} {Active matter at the interface between materials science and cell biology},\ }\href@noop {} {\bibfield  {journal} {\bibinfo  {journal} {Nature Reviews Materials}\ }\textbf {\bibinfo {volume} {2}},\ \bibinfo {pages} {1} (\bibinfo {year} {2017})}\BibitemShut {NoStop}%
\bibitem [{\citenamefont {Dulaney}\ \emph {et~al.}(2021)\citenamefont {Dulaney}, \citenamefont {Mallory},\ and\ \citenamefont {Brady}}]{Dulaney2021-kf}%
  \BibitemOpen
  \bibfield  {author} {\bibinfo {author} {\bibfnamefont {A.~R.}\ \bibnamefont {Dulaney}}, \bibinfo {author} {\bibfnamefont {S.~A.}\ \bibnamefont {Mallory}},\ and\ \bibinfo {author} {\bibfnamefont {J.~F.}\ \bibnamefont {Brady}},\ }\bibfield  {title} {\bibinfo {title} {The ``isothermal'' compressibility of active matter},\ }\href@noop {} {\bibfield  {journal} {\bibinfo  {journal} {J. Chem. Phys.}\ }\textbf {\bibinfo {volume} {154}},\ \bibinfo {pages} {014902} (\bibinfo {year} {2021})}\BibitemShut {NoStop}%
\bibitem [{\citenamefont {Choi}\ \emph {et~al.}(2025)\citenamefont {Choi}, \citenamefont {Schiltz-Rouse}, \citenamefont {Bayati},\ and\ \citenamefont {Mallory}}]{Choi2025-mu}%
  \BibitemOpen
  \bibfield  {author} {\bibinfo {author} {\bibfnamefont {Y.}~\bibnamefont {Choi}}, \bibinfo {author} {\bibfnamefont {E.}~\bibnamefont {Schiltz-Rouse}}, \bibinfo {author} {\bibfnamefont {P.}~\bibnamefont {Bayati}},\ and\ \bibinfo {author} {\bibfnamefont {S.~A.}\ \bibnamefont {Mallory}},\ }\bibfield  {title} {\bibinfo {title} {Sedimentation equilibrium as a probe of the pressure equation of state of active colloids},\ }\href@noop {} {\bibfield  {journal} {\bibinfo  {journal} {Soft Matter}\ }\textbf {\bibinfo {volume} {21}},\ \bibinfo {pages} {7449} (\bibinfo {year} {2025})}\BibitemShut {NoStop}%
\bibitem [{\citenamefont {Ghosh}\ \emph {et~al.}(2015)\citenamefont {Ghosh}, \citenamefont {Li}, \citenamefont {Marchegiani},\ and\ \citenamefont {Marchesoni}}]{Ghosh2015-xp}%
  \BibitemOpen
  \bibfield  {author} {\bibinfo {author} {\bibfnamefont {P.~K.}\ \bibnamefont {Ghosh}}, \bibinfo {author} {\bibfnamefont {Y.}~\bibnamefont {Li}}, \bibinfo {author} {\bibfnamefont {G.}~\bibnamefont {Marchegiani}},\ and\ \bibinfo {author} {\bibfnamefont {F.}~\bibnamefont {Marchesoni}},\ }\bibfield  {title} {\bibinfo {title} {Communication: Memory effects and active brownian diffusion},\ }\href@noop {} {\bibfield  {journal} {\bibinfo  {journal} {J. Chem. Phys.}\ }\textbf {\bibinfo {volume} {143}},\ \bibinfo {pages} {211101} (\bibinfo {year} {2015})}\BibitemShut {NoStop}%
\bibitem [{\citenamefont {Howse}\ \emph {et~al.}(2007)\citenamefont {Howse}, \citenamefont {Jones}, \citenamefont {Ryan}, \citenamefont {Gough}, \citenamefont {Vafabakhsh},\ and\ \citenamefont {Golestanian}}]{Howse2007-vh}%
  \BibitemOpen
  \bibfield  {author} {\bibinfo {author} {\bibfnamefont {J.~R.}\ \bibnamefont {Howse}}, \bibinfo {author} {\bibfnamefont {R.~A.~L.}\ \bibnamefont {Jones}}, \bibinfo {author} {\bibfnamefont {A.~J.}\ \bibnamefont {Ryan}}, \bibinfo {author} {\bibfnamefont {T.}~\bibnamefont {Gough}}, \bibinfo {author} {\bibfnamefont {R.}~\bibnamefont {Vafabakhsh}},\ and\ \bibinfo {author} {\bibfnamefont {R.}~\bibnamefont {Golestanian}},\ }\bibfield  {title} {\bibinfo {title} {Self-motile colloidal particles: from directed propulsion to random walk},\ }\href@noop {} {\bibfield  {journal} {\bibinfo  {journal} {Phys. Rev. Lett.}\ }\textbf {\bibinfo {volume} {99}},\ \bibinfo {pages} {048102} (\bibinfo {year} {2007})}\BibitemShut {NoStop}%
\bibitem [{\citenamefont {Sevilla}\ and\ \citenamefont {Gómez~Nava}(2014)}]{Sevilla2014-yt}%
  \BibitemOpen
  \bibfield  {author} {\bibinfo {author} {\bibfnamefont {F.~J.}\ \bibnamefont {Sevilla}}\ and\ \bibinfo {author} {\bibfnamefont {L.~A.}\ \bibnamefont {Gómez~Nava}},\ }\bibfield  {title} {\bibinfo {title} {Theory of diffusion of active particles that move at constant speed in two dimensions},\ }\href@noop {} {\bibfield  {journal} {\bibinfo  {journal} {Phys. Rev. E Stat. Nonlin. Soft Matter Phys.}\ }\textbf {\bibinfo {volume} {90}},\ \bibinfo {pages} {022130} (\bibinfo {year} {2014})}\BibitemShut {NoStop}%
\bibitem [{\citenamefont {Sevilla}\ and\ \citenamefont {Sandoval}(2015)}]{Sevilla2015-jn}%
  \BibitemOpen
  \bibfield  {author} {\bibinfo {author} {\bibfnamefont {F.~J.}\ \bibnamefont {Sevilla}}\ and\ \bibinfo {author} {\bibfnamefont {M.}~\bibnamefont {Sandoval}},\ }\bibfield  {title} {\bibinfo {title} {Smoluchowski diffusion equation for active brownian swimmers},\ }\href@noop {} {\bibfield  {journal} {\bibinfo  {journal} {Phys. Rev. E Stat. Nonlin. Soft Matter Phys.}\ }\textbf {\bibinfo {volume} {91}},\ \bibinfo {pages} {052150} (\bibinfo {year} {2015})}\BibitemShut {NoStop}%
\bibitem [{\citenamefont {Basu}\ \emph {et~al.}(2018)\citenamefont {Basu}, \citenamefont {Majumdar}, \citenamefont {Rosso},\ and\ \citenamefont {Schehr}}]{Basu2018-th}%
  \BibitemOpen
  \bibfield  {author} {\bibinfo {author} {\bibfnamefont {U.}~\bibnamefont {Basu}}, \bibinfo {author} {\bibfnamefont {S.~N.}\ \bibnamefont {Majumdar}}, \bibinfo {author} {\bibfnamefont {A.}~\bibnamefont {Rosso}},\ and\ \bibinfo {author} {\bibfnamefont {G.}~\bibnamefont {Schehr}},\ }\bibfield  {title} {\bibinfo {title} {Active brownian motion in two dimensions},\ }\href@noop {} {\bibfield  {journal} {\bibinfo  {journal} {Phys. Rev. E}\ }\textbf {\bibinfo {volume} {98}},\ \bibinfo {pages} {062121} (\bibinfo {year} {2018})}\BibitemShut {NoStop}%
\bibitem [{\citenamefont {Dulaney}\ and\ \citenamefont {Brady}(2020)}]{Dulaney2020-mb}%
  \BibitemOpen
  \bibfield  {author} {\bibinfo {author} {\bibfnamefont {A.~R.}\ \bibnamefont {Dulaney}}\ and\ \bibinfo {author} {\bibfnamefont {J.~F.}\ \bibnamefont {Brady}},\ }\bibfield  {title} {\bibinfo {title} {Waves in active matter: The transition from ballistic to diffusive behavior},\ }\href@noop {} {\bibfield  {journal} {\bibinfo  {journal} {Phys Rev E}\ }\textbf {\bibinfo {volume} {101}},\ \bibinfo {pages} {052609} (\bibinfo {year} {2020})}\BibitemShut {NoStop}%
\bibitem [{\citenamefont {Mallory}\ \emph {et~al.}(2020)\citenamefont {Mallory}, \citenamefont {Bowers},\ and\ \citenamefont {Cacciuto}}]{Mallory2020-tq}%
  \BibitemOpen
  \bibfield  {author} {\bibinfo {author} {\bibfnamefont {S.~A.}\ \bibnamefont {Mallory}}, \bibinfo {author} {\bibfnamefont {M.~L.}\ \bibnamefont {Bowers}},\ and\ \bibinfo {author} {\bibfnamefont {A.}~\bibnamefont {Cacciuto}},\ }\bibfield  {title} {\bibinfo {title} {Universal reshaping of arrested colloidal gels via active doping},\ }\href@noop {} {\bibfield  {journal} {\bibinfo  {journal} {J. Chem. Phys.}\ }\textbf {\bibinfo {volume} {153}},\ \bibinfo {pages} {084901} (\bibinfo {year} {2020})}\BibitemShut {NoStop}%
\bibitem [{\citenamefont {Sandoval}(2020)}]{Sandoval2020-yy}%
  \BibitemOpen
  \bibfield  {author} {\bibinfo {author} {\bibfnamefont {M.}~\bibnamefont {Sandoval}},\ }\bibfield  {title} {\bibinfo {title} {Pressure and diffusion of active matter with inertia},\ }\href@noop {} {\bibfield  {journal} {\bibinfo  {journal} {Phys Rev E}\ }\textbf {\bibinfo {volume} {101}},\ \bibinfo {pages} {012606} (\bibinfo {year} {2020})}\BibitemShut {NoStop}%
\bibitem [{\citenamefont {Schakenraad}\ \emph {et~al.}(2020)\citenamefont {Schakenraad}, \citenamefont {Ravazzano}, \citenamefont {Sarkar}, \citenamefont {Wondergem}, \citenamefont {Merks},\ and\ \citenamefont {Giomi}}]{Schakenraad2020-dg}%
  \BibitemOpen
  \bibfield  {author} {\bibinfo {author} {\bibfnamefont {K.}~\bibnamefont {Schakenraad}}, \bibinfo {author} {\bibfnamefont {L.}~\bibnamefont {Ravazzano}}, \bibinfo {author} {\bibfnamefont {N.}~\bibnamefont {Sarkar}}, \bibinfo {author} {\bibfnamefont {J.~A.~J.}\ \bibnamefont {Wondergem}}, \bibinfo {author} {\bibfnamefont {R.~M.~H.}\ \bibnamefont {Merks}},\ and\ \bibinfo {author} {\bibfnamefont {L.}~\bibnamefont {Giomi}},\ }\bibfield  {title} {\bibinfo {title} {Topotaxis of active brownian particles},\ }\href@noop {} {\bibfield  {journal} {\bibinfo  {journal} {Phys Rev E}\ }\textbf {\bibinfo {volume} {101}},\ \bibinfo {pages} {032602} (\bibinfo {year} {2020})}\BibitemShut {NoStop}%
\bibitem [{\citenamefont {Caraglio}\ and\ \citenamefont {Franosch}(2022)}]{Caraglio2022-dz}%
  \BibitemOpen
  \bibfield  {author} {\bibinfo {author} {\bibfnamefont {M.}~\bibnamefont {Caraglio}}\ and\ \bibinfo {author} {\bibfnamefont {T.}~\bibnamefont {Franosch}},\ }\bibfield  {title} {\bibinfo {title} {Analytic solution of an active brownian particle in a harmonic well},\ }\href@noop {} {\bibfield  {journal} {\bibinfo  {journal} {Phys. Rev. Lett.}\ }\textbf {\bibinfo {volume} {129}},\ \bibinfo {pages} {158001} (\bibinfo {year} {2022})}\BibitemShut {NoStop}%
\bibitem [{\citenamefont {Modica}\ \emph {et~al.}(2023)\citenamefont {Modica}, \citenamefont {Omar},\ and\ \citenamefont {Takatori}}]{Modica2023-tb}%
  \BibitemOpen
  \bibfield  {author} {\bibinfo {author} {\bibfnamefont {K.~J.}\ \bibnamefont {Modica}}, \bibinfo {author} {\bibfnamefont {A.~K.}\ \bibnamefont {Omar}},\ and\ \bibinfo {author} {\bibfnamefont {S.~C.}\ \bibnamefont {Takatori}},\ }\bibfield  {title} {\bibinfo {title} {Boundary design regulates the diffusion of active matter in heterogeneous environments},\ }\href@noop {} {\bibfield  {journal} {\bibinfo  {journal} {Soft Matter}\ }\textbf {\bibinfo {volume} {19}},\ \bibinfo {pages} {1890} (\bibinfo {year} {2023})}\BibitemShut {NoStop}%
\bibitem [{\citenamefont {Bayati}\ and\ \citenamefont {Mallory}(2024)}]{Bayati2024-kd}%
  \BibitemOpen
  \bibfield  {author} {\bibinfo {author} {\bibfnamefont {P.}~\bibnamefont {Bayati}}\ and\ \bibinfo {author} {\bibfnamefont {S.~A.}\ \bibnamefont {Mallory}},\ }\bibfield  {title} {\bibinfo {title} {Orbits, spirals, and trapped states: Dynamics of a phoretic janus particle in a radial concentration gradient},\ }\href@noop {} {\bibfield  {journal} {\bibinfo  {journal} {ACS Nano}\ }\textbf {\bibinfo {volume} {18}},\ \bibinfo {pages} {23047} (\bibinfo {year} {2024})}\BibitemShut {NoStop}%
\bibitem [{\citenamefont {Bayati}\ and\ \citenamefont {Mallory}(2025)}]{Bayati2025-kk}%
  \BibitemOpen
  \bibfield  {author} {\bibinfo {author} {\bibfnamefont {P.}~\bibnamefont {Bayati}}\ and\ \bibinfo {author} {\bibfnamefont {S.~A.}\ \bibnamefont {Mallory}},\ }\bibfield  {title} {\bibinfo {title} {Inferring surface slip in active colloids from flow fields using physics-informed neural networks},\ }\href@noop {} {\bibfield  {journal} {\bibinfo  {journal} {arXiv [cond-mat.soft]}\ } (\bibinfo {year} {2025})}\BibitemShut {NoStop}%
\bibitem [{\citenamefont {Soto}(2025)}]{Soto2025-lg}%
  \BibitemOpen
  \bibfield  {author} {\bibinfo {author} {\bibfnamefont {R.}~\bibnamefont {Soto}},\ }\bibfield  {title} {\bibinfo {title} {Self-diffusive dynamics of active brownian particles at moderate densities},\ }\href@noop {} {\bibfield  {journal} {\bibinfo  {journal} {Phys. Fluids (1994)}\ }\textbf {\bibinfo {volume} {37}},\ \bibinfo {pages} {033309} (\bibinfo {year} {2025})}\BibitemShut {NoStop}%
\bibitem [{\citenamefont {J{\"u}licher}\ \emph {et~al.}(2018)\citenamefont {J{\"u}licher}, \citenamefont {Grill},\ and\ \citenamefont {Salbreux}}]{julicher2018hydrodynamic}%
  \BibitemOpen
  \bibfield  {author} {\bibinfo {author} {\bibfnamefont {F.}~\bibnamefont {J{\"u}licher}}, \bibinfo {author} {\bibfnamefont {S.~W.}\ \bibnamefont {Grill}},\ and\ \bibinfo {author} {\bibfnamefont {G.}~\bibnamefont {Salbreux}},\ }\bibfield  {title} {\bibinfo {title} {Hydrodynamic theory of active matter},\ }\href@noop {} {\bibfield  {journal} {\bibinfo  {journal} {Reports on Progress in Physics}\ }\textbf {\bibinfo {volume} {81}},\ \bibinfo {pages} {076601} (\bibinfo {year} {2018})}\BibitemShut {NoStop}%
\bibitem [{\citenamefont {Te~Vrugt}\ and\ \citenamefont {Wittkowski}(2025)}]{te2025metareview}%
  \BibitemOpen
  \bibfield  {author} {\bibinfo {author} {\bibfnamefont {M.}~\bibnamefont {Te~Vrugt}}\ and\ \bibinfo {author} {\bibfnamefont {R.}~\bibnamefont {Wittkowski}},\ }\bibfield  {title} {\bibinfo {title} {Metareview: a survey of active matter reviews},\ }\href@noop {} {\bibfield  {journal} {\bibinfo  {journal} {The European Physical Journal E}\ }\textbf {\bibinfo {volume} {48}},\ \bibinfo {pages} {12} (\bibinfo {year} {2025})}\BibitemShut {NoStop}%
\bibitem [{\citenamefont {Ramaswamy}(2019)}]{ramaswamy2019active}%
  \BibitemOpen
  \bibfield  {author} {\bibinfo {author} {\bibfnamefont {S.}~\bibnamefont {Ramaswamy}},\ }\bibfield  {title} {\bibinfo {title} {Active fluids},\ }\href@noop {} {\bibfield  {journal} {\bibinfo  {journal} {Nature Reviews Physics}\ }\textbf {\bibinfo {volume} {1}},\ \bibinfo {pages} {640} (\bibinfo {year} {2019})}\BibitemShut {NoStop}%
\bibitem [{\citenamefont {Schiltz-Rouse}\ \emph {et~al.}(2023)\citenamefont {Schiltz-Rouse}, \citenamefont {Row},\ and\ \citenamefont {Mallory}}]{Schiltz-Rouse2023-vj}%
  \BibitemOpen
  \bibfield  {author} {\bibinfo {author} {\bibfnamefont {E.}~\bibnamefont {Schiltz-Rouse}}, \bibinfo {author} {\bibfnamefont {H.}~\bibnamefont {Row}},\ and\ \bibinfo {author} {\bibfnamefont {S.~A.}\ \bibnamefont {Mallory}},\ }\bibfield  {title} {\bibinfo {title} {Kinetic temperature and pressure of an active tonks gas},\ }\href@noop {} {\bibfield  {journal} {\bibinfo  {journal} {Phys. Rev. E}\ }\textbf {\bibinfo {volume} {108}},\ \bibinfo {pages} {064601} (\bibinfo {year} {2023})}\BibitemShut {NoStop}%
\bibitem [{\citenamefont {Akintunde}\ \emph {et~al.}(2025)\citenamefont {Akintunde}, \citenamefont {Bayati}, \citenamefont {Row},\ and\ \citenamefont {Mallory}}]{akintunde2025single}%
  \BibitemOpen
  \bibfield  {author} {\bibinfo {author} {\bibfnamefont {A.}~\bibnamefont {Akintunde}}, \bibinfo {author} {\bibfnamefont {P.}~\bibnamefont {Bayati}}, \bibinfo {author} {\bibfnamefont {H.}~\bibnamefont {Row}},\ and\ \bibinfo {author} {\bibfnamefont {S.~A.}\ \bibnamefont {Mallory}},\ }\bibfield  {title} {\bibinfo {title} {Single-file diffusion of active brownian particles},\ }\href@noop {} {\bibfield  {journal} {\bibinfo  {journal} {The Journal of Chemical Physics}\ }\textbf {\bibinfo {volume} {162}} (\bibinfo {year} {2025})}\BibitemShut {NoStop}%
\bibitem [{\citenamefont {Omar}\ \emph {et~al.}(2023)\citenamefont {Omar}, \citenamefont {Klymko}, \citenamefont {GrandPre}, \citenamefont {Geissler},\ and\ \citenamefont {Brady}}]{Omar2023-lq}%
  \BibitemOpen
  \bibfield  {author} {\bibinfo {author} {\bibfnamefont {A.~K.}\ \bibnamefont {Omar}}, \bibinfo {author} {\bibfnamefont {K.}~\bibnamefont {Klymko}}, \bibinfo {author} {\bibfnamefont {T.}~\bibnamefont {GrandPre}}, \bibinfo {author} {\bibfnamefont {P.~L.}\ \bibnamefont {Geissler}},\ and\ \bibinfo {author} {\bibfnamefont {J.~F.}\ \bibnamefont {Brady}},\ }\bibfield  {title} {\bibinfo {title} {Tuning nonequilibrium phase transitions with inertia},\ }\href@noop {} {\bibfield  {journal} {\bibinfo  {journal} {J. Chem. Phys.}\ }\textbf {\bibinfo {volume} {158}},\ \bibinfo {pages} {074904} (\bibinfo {year} {2023})}\BibitemShut {NoStop}%
\bibitem [{\citenamefont {Shaebani}\ \emph {et~al.}(2020)\citenamefont {Shaebani}, \citenamefont {Wysocki}, \citenamefont {Winkler}, \citenamefont {Gompper},\ and\ \citenamefont {Rieger}}]{Shaebani2020-yf}%
  \BibitemOpen
  \bibfield  {author} {\bibinfo {author} {\bibfnamefont {M.~R.}\ \bibnamefont {Shaebani}}, \bibinfo {author} {\bibfnamefont {A.}~\bibnamefont {Wysocki}}, \bibinfo {author} {\bibfnamefont {R.~G.}\ \bibnamefont {Winkler}}, \bibinfo {author} {\bibfnamefont {G.}~\bibnamefont {Gompper}},\ and\ \bibinfo {author} {\bibfnamefont {H.}~\bibnamefont {Rieger}},\ }\bibfield  {title} {\bibinfo {title} {Computational models for active matter},\ }\href@noop {} {\bibfield  {journal} {\bibinfo  {journal} {Nat. Rev. Phys.}\ }\textbf {\bibinfo {volume} {2}},\ \bibinfo {pages} {181} (\bibinfo {year} {2020})}\BibitemShut {NoStop}%
\bibitem [{\citenamefont {Bialké}\ \emph {et~al.}(2013)\citenamefont {Bialké}, \citenamefont {Löwen},\ and\ \citenamefont {Speck}}]{Bialke2013-gy}%
  \BibitemOpen
  \bibfield  {author} {\bibinfo {author} {\bibfnamefont {J.}~\bibnamefont {Bialké}}, \bibinfo {author} {\bibfnamefont {H.}~\bibnamefont {Löwen}},\ and\ \bibinfo {author} {\bibfnamefont {T.}~\bibnamefont {Speck}},\ }\bibfield  {title} {\bibinfo {title} {Microscopic theory for the phase separation of self-propelled repulsive disks},\ }\href@noop {} {\bibfield  {journal} {\bibinfo  {journal} {EPL}\ }\textbf {\bibinfo {volume} {103}},\ \bibinfo {pages} {30008} (\bibinfo {year} {2013})}\BibitemShut {NoStop}%
\bibitem [{\citenamefont {Speck}\ \emph {et~al.}(2015)\citenamefont {Speck}, \citenamefont {Menzel}, \citenamefont {Bialké},\ and\ \citenamefont {Löwen}}]{Speck2015-df}%
  \BibitemOpen
  \bibfield  {author} {\bibinfo {author} {\bibfnamefont {T.}~\bibnamefont {Speck}}, \bibinfo {author} {\bibfnamefont {A.~M.}\ \bibnamefont {Menzel}}, \bibinfo {author} {\bibfnamefont {J.}~\bibnamefont {Bialké}},\ and\ \bibinfo {author} {\bibfnamefont {H.}~\bibnamefont {Löwen}},\ }\bibfield  {title} {\bibinfo {title} {Dynamical mean-field theory and weakly non-linear analysis for the phase separation of active brownian particles},\ }\href@noop {} {\bibfield  {journal} {\bibinfo  {journal} {J. Chem. Phys.}\ }\textbf {\bibinfo {volume} {142}},\ \bibinfo {pages} {224109} (\bibinfo {year} {2015})}\BibitemShut {NoStop}%
\bibitem [{\citenamefont {Wittkowski}\ \emph {et~al.}(2017)\citenamefont {Wittkowski}, \citenamefont {Stenhammar},\ and\ \citenamefont {Cates}}]{Wittkowski2017-ty}%
  \BibitemOpen
  \bibfield  {author} {\bibinfo {author} {\bibfnamefont {R.}~\bibnamefont {Wittkowski}}, \bibinfo {author} {\bibfnamefont {J.}~\bibnamefont {Stenhammar}},\ and\ \bibinfo {author} {\bibfnamefont {M.~E.}\ \bibnamefont {Cates}},\ }\bibfield  {title} {\bibinfo {title} {Nonequilibrium dynamics of mixtures of active and passive colloidal particles},\ }\href@noop {} {\bibfield  {journal} {\bibinfo  {journal} {New J. Phys.}\ }\textbf {\bibinfo {volume} {19}},\ \bibinfo {pages} {105003} (\bibinfo {year} {2017})}\BibitemShut {NoStop}%
\bibitem [{\citenamefont {Jeggle}\ \emph {et~al.}(2020)\citenamefont {Jeggle}, \citenamefont {Stenhammar},\ and\ \citenamefont {Wittkowski}}]{Jeggle2020-ca}%
  \BibitemOpen
  \bibfield  {author} {\bibinfo {author} {\bibfnamefont {J.}~\bibnamefont {Jeggle}}, \bibinfo {author} {\bibfnamefont {J.}~\bibnamefont {Stenhammar}},\ and\ \bibinfo {author} {\bibfnamefont {R.}~\bibnamefont {Wittkowski}},\ }\bibfield  {title} {\bibinfo {title} {Pair-distribution function of active brownian spheres in two spatial dimensions: Simulation results and analytic representation},\ }\href@noop {} {\bibfield  {journal} {\bibinfo  {journal} {J. Chem. Phys.}\ }\textbf {\bibinfo {volume} {152}},\ \bibinfo {pages} {194903} (\bibinfo {year} {2020})}\BibitemShut {NoStop}%
\bibitem [{\citenamefont {Großmann}\ \emph {et~al.}(2020)\citenamefont {Großmann}, \citenamefont {Aranson},\ and\ \citenamefont {Peruani}}]{Grossmann2020-th}%
  \BibitemOpen
  \bibfield  {author} {\bibinfo {author} {\bibfnamefont {R.}~\bibnamefont {Großmann}}, \bibinfo {author} {\bibfnamefont {I.~S.}\ \bibnamefont {Aranson}},\ and\ \bibinfo {author} {\bibfnamefont {F.}~\bibnamefont {Peruani}},\ }\bibfield  {title} {\bibinfo {title} {A particle-field approach bridges phase separation and collective motion in active matter},\ }\href@noop {} {\bibfield  {journal} {\bibinfo  {journal} {Nat. Commun.}\ }\textbf {\bibinfo {volume} {11}},\ \bibinfo {pages} {5365} (\bibinfo {year} {2020})}\BibitemShut {NoStop}%
\bibitem [{\citenamefont {Dasgupta}\ \emph {et~al.}(2026)\citenamefont {Dasgupta}, \citenamefont {Mandal}, \citenamefont {K~Mukhopadhyay},\ and\ \citenamefont {Liebchen}}]{Dasgupta2026-cz}%
  \BibitemOpen
  \bibfield  {author} {\bibinfo {author} {\bibfnamefont {W.}~\bibnamefont {Dasgupta}}, \bibinfo {author} {\bibfnamefont {S.}~\bibnamefont {Mandal}}, \bibinfo {author} {\bibfnamefont {A.}~\bibnamefont {K~Mukhopadhyay}},\ and\ \bibinfo {author} {\bibfnamefont {B.}~\bibnamefont {Liebchen}},\ }\bibfield  {title} {\bibinfo {title} {Learning microstructure in active matter},\ }\href@noop {} {\bibfield  {journal} {\bibinfo  {journal} {arXiv [cond-mat.soft]}\ } (\bibinfo {year} {2026})}\BibitemShut {NoStop}%
\bibitem [{\citenamefont {Levis}\ \emph {et~al.}(2017)\citenamefont {Levis}, \citenamefont {Codina},\ and\ \citenamefont {Pagonabarraga}}]{Levis2017-je}%
  \BibitemOpen
  \bibfield  {author} {\bibinfo {author} {\bibfnamefont {D.}~\bibnamefont {Levis}}, \bibinfo {author} {\bibfnamefont {J.}~\bibnamefont {Codina}},\ and\ \bibinfo {author} {\bibfnamefont {I.}~\bibnamefont {Pagonabarraga}},\ }\bibfield  {title} {\bibinfo {title} {Active brownian equation of state: metastability and phase coexistence},\ }\href@noop {} {\bibfield  {journal} {\bibinfo  {journal} {Soft Matter}\ }\textbf {\bibinfo {volume} {13}},\ \bibinfo {pages} {8113} (\bibinfo {year} {2017})}\BibitemShut {NoStop}%
\bibitem [{\citenamefont {Lee}\ and\ \citenamefont {Brenner}(2016)}]{lee2016molecular}%
  \BibitemOpen
  \bibfield  {author} {\bibinfo {author} {\bibfnamefont {L.}~\bibnamefont {Lee}}\ and\ \bibinfo {author} {\bibfnamefont {H.}~\bibnamefont {Brenner}},\ }\href@noop {} {\emph {\bibinfo {title} {Molecular Thermodynamics of Nonideal Fluids}}}\ (\bibinfo  {publisher} {Butterworth-Heinemann},\ \bibinfo {year} {2016})\BibitemShut {NoStop}%
\bibitem [{\citenamefont {Fily}\ and\ \citenamefont {Marchetti}(2012)}]{Fily2012-ej}%
  \BibitemOpen
  \bibfield  {author} {\bibinfo {author} {\bibfnamefont {Y.}~\bibnamefont {Fily}}\ and\ \bibinfo {author} {\bibfnamefont {M.~C.}\ \bibnamefont {Marchetti}},\ }\bibfield  {title} {\bibinfo {title} {Athermal phase separation of self-propelled particles with no alignment},\ }\href@noop {} {\bibfield  {journal} {\bibinfo  {journal} {Phys. Rev. Lett.}\ }\textbf {\bibinfo {volume} {108}},\ \bibinfo {pages} {235702} (\bibinfo {year} {2012})}\BibitemShut {NoStop}%
\bibitem [{\citenamefont {Fily}\ \emph {et~al.}(2014)\citenamefont {Fily}, \citenamefont {Henkes},\ and\ \citenamefont {Marchetti}}]{Fily2014-vf}%
  \BibitemOpen
  \bibfield  {author} {\bibinfo {author} {\bibfnamefont {Y.}~\bibnamefont {Fily}}, \bibinfo {author} {\bibfnamefont {S.}~\bibnamefont {Henkes}},\ and\ \bibinfo {author} {\bibfnamefont {M.~C.}\ \bibnamefont {Marchetti}},\ }\bibfield  {title} {\bibinfo {title} {Freezing and phase separation of self-propelled disks},\ }\href@noop {} {\bibfield  {journal} {\bibinfo  {journal} {Soft Matter}\ }\textbf {\bibinfo {volume} {10}},\ \bibinfo {pages} {2132} (\bibinfo {year} {2014})}\BibitemShut {NoStop}%
\bibitem [{\citenamefont {Levis}\ and\ \citenamefont {Berthier}(2014)}]{Levis2014-fe}%
  \BibitemOpen
  \bibfield  {author} {\bibinfo {author} {\bibfnamefont {D.}~\bibnamefont {Levis}}\ and\ \bibinfo {author} {\bibfnamefont {L.}~\bibnamefont {Berthier}},\ }\bibfield  {title} {\bibinfo {title} {Clustering and heterogeneous dynamics in a kinetic monte carlo model of self-propelled hard disks},\ }\href@noop {} {\bibfield  {journal} {\bibinfo  {journal} {Phys. Rev. E Stat. Nonlin. Soft Matter Phys.}\ }\textbf {\bibinfo {volume} {89}},\ \bibinfo {pages} {062301} (\bibinfo {year} {2014})}\BibitemShut {NoStop}%
\bibitem [{\citenamefont {Digregorio}\ \emph {et~al.}(2018)\citenamefont {Digregorio}, \citenamefont {Levis}, \citenamefont {Suma}, \citenamefont {Cugliandolo}, \citenamefont {Gonnella},\ and\ \citenamefont {Pagonabarraga}}]{Digregorio2018-ve}%
  \BibitemOpen
  \bibfield  {author} {\bibinfo {author} {\bibfnamefont {P.}~\bibnamefont {Digregorio}}, \bibinfo {author} {\bibfnamefont {D.}~\bibnamefont {Levis}}, \bibinfo {author} {\bibfnamefont {A.}~\bibnamefont {Suma}}, \bibinfo {author} {\bibfnamefont {L.~F.}\ \bibnamefont {Cugliandolo}}, \bibinfo {author} {\bibfnamefont {G.}~\bibnamefont {Gonnella}},\ and\ \bibinfo {author} {\bibfnamefont {I.}~\bibnamefont {Pagonabarraga}},\ }\bibfield  {title} {\bibinfo {title} {Full phase diagram of active brownian disks: From melting to motility-induced phase separation},\ }\href@noop {} {\bibfield  {journal} {\bibinfo  {journal} {Phys. Rev. Lett.}\ }\textbf {\bibinfo {volume} {121}},\ \bibinfo {pages} {098003} (\bibinfo {year} {2018})}\BibitemShut {NoStop}%
\bibitem [{\citenamefont {Mallory}\ \emph {et~al.}(2021)\citenamefont {Mallory}, \citenamefont {Omar},\ and\ \citenamefont {Brady}}]{Mallory2021-hi}%
  \BibitemOpen
  \bibfield  {author} {\bibinfo {author} {\bibfnamefont {S.~A.}\ \bibnamefont {Mallory}}, \bibinfo {author} {\bibfnamefont {A.~K.}\ \bibnamefont {Omar}},\ and\ \bibinfo {author} {\bibfnamefont {J.~F.}\ \bibnamefont {Brady}},\ }\bibfield  {title} {\bibinfo {title} {Dynamic overlap concentration scale of active colloids},\ }\href@noop {} {\bibfield  {journal} {\bibinfo  {journal} {Phys Rev E}\ }\textbf {\bibinfo {volume} {104}},\ \bibinfo {pages} {044612} (\bibinfo {year} {2021})}\BibitemShut {NoStop}%
\bibitem [{\citenamefont {Bernard}\ and\ \citenamefont {Krauth}(2011)}]{Bernard2011-qm}%
  \BibitemOpen
  \bibfield  {author} {\bibinfo {author} {\bibfnamefont {E.~P.}\ \bibnamefont {Bernard}}\ and\ \bibinfo {author} {\bibfnamefont {W.}~\bibnamefont {Krauth}},\ }\bibfield  {title} {\bibinfo {title} {Two-step melting in two dimensions: first-order liquid-hexatic transition},\ }\href@noop {} {\bibfield  {journal} {\bibinfo  {journal} {Phys. Rev. Lett.}\ }\textbf {\bibinfo {volume} {107}},\ \bibinfo {pages} {155704} (\bibinfo {year} {2011})}\BibitemShut {NoStop}%
\bibitem [{\citenamefont {ten Hagen}\ \emph {et~al.}(2011)\citenamefont {ten Hagen}, \citenamefont {van Teeffelen},\ and\ \citenamefont {L{\"o}wen}}]{ten2011brownian}%
  \BibitemOpen
  \bibfield  {author} {\bibinfo {author} {\bibfnamefont {B.}~\bibnamefont {ten Hagen}}, \bibinfo {author} {\bibfnamefont {S.}~\bibnamefont {van Teeffelen}},\ and\ \bibinfo {author} {\bibfnamefont {H.}~\bibnamefont {L{\"o}wen}},\ }\bibfield  {title} {\bibinfo {title} {Brownian motion of a self-propelled particle},\ }\href@noop {} {\bibfield  {journal} {\bibinfo  {journal} {Journal of Physics: Condensed Matter}\ }\textbf {\bibinfo {volume} {23}},\ \bibinfo {pages} {194119} (\bibinfo {year} {2011})}\BibitemShut {NoStop}%
\bibitem [{\citenamefont {Stopper}\ \emph {et~al.}(2018)\citenamefont {Stopper}, \citenamefont {Thorneywork}, \citenamefont {Dullens},\ and\ \citenamefont {Roth}}]{Stopper2018-nk}%
  \BibitemOpen
  \bibfield  {author} {\bibinfo {author} {\bibfnamefont {D.}~\bibnamefont {Stopper}}, \bibinfo {author} {\bibfnamefont {A.~L.}\ \bibnamefont {Thorneywork}}, \bibinfo {author} {\bibfnamefont {R.~P.~A.}\ \bibnamefont {Dullens}},\ and\ \bibinfo {author} {\bibfnamefont {R.}~\bibnamefont {Roth}},\ }\bibfield  {title} {\bibinfo {title} {Bulk dynamics of brownian hard disks: Dynamical density functional theory versus experiments on two-dimensional colloidal hard spheres},\ }\href@noop {} {\bibfield  {journal} {\bibinfo  {journal} {J. Chem. Phys.}\ }\textbf {\bibinfo {volume} {148}},\ \bibinfo {pages} {104501} (\bibinfo {year} {2018})}\BibitemShut {NoStop}%
\bibitem [{Note1()}]{Note1}%
  \BibitemOpen
  \bibinfo {note} {See Supplemental Material at [URL] for details.}\BibitemShut {Stop}%
\end{thebibliography}

%apsrev4-2.bst 2019-01-14 (MD) hand-edited version of apsrev4-1.bst
%Control: key (0)
%Control: author (8) initials jnrlst
%Control: editor formatted (1) identically to author
%Control: production of article title (0) allowed
%Control: page (0) single
%Control: year (1) truncated
%Control: production of eprint (0) enabled
%

\end{document}

\preprint{APS/123-QED}
\title{Supplemental Material: \\ Structure and Memory Control Self-Diffusion in Active Matter}

\author{Akinlade Akintunde}
\affiliation{Department of Chemistry, The Pennsylvania State University, University Park, Pennsylvania, 16802, USA}

\author{Stewart A. Mallory}
\email{sam7808@psu.edu}
\affiliation{Department of Chemistry, The Pennsylvania State University, University Park, Pennsylvania, 16802, USA}
\affiliation{Department of Chemical Engineering, The Pennsylvania State University, University Park, Pennsylvania, 16802, USA}

\date{January 2025}
\maketitle

\section{Anisotropic PAIR CORRELATION FUNCTION}
\begin{figure}[h!]
\centering
\includegraphics[width=.85\textwidth]{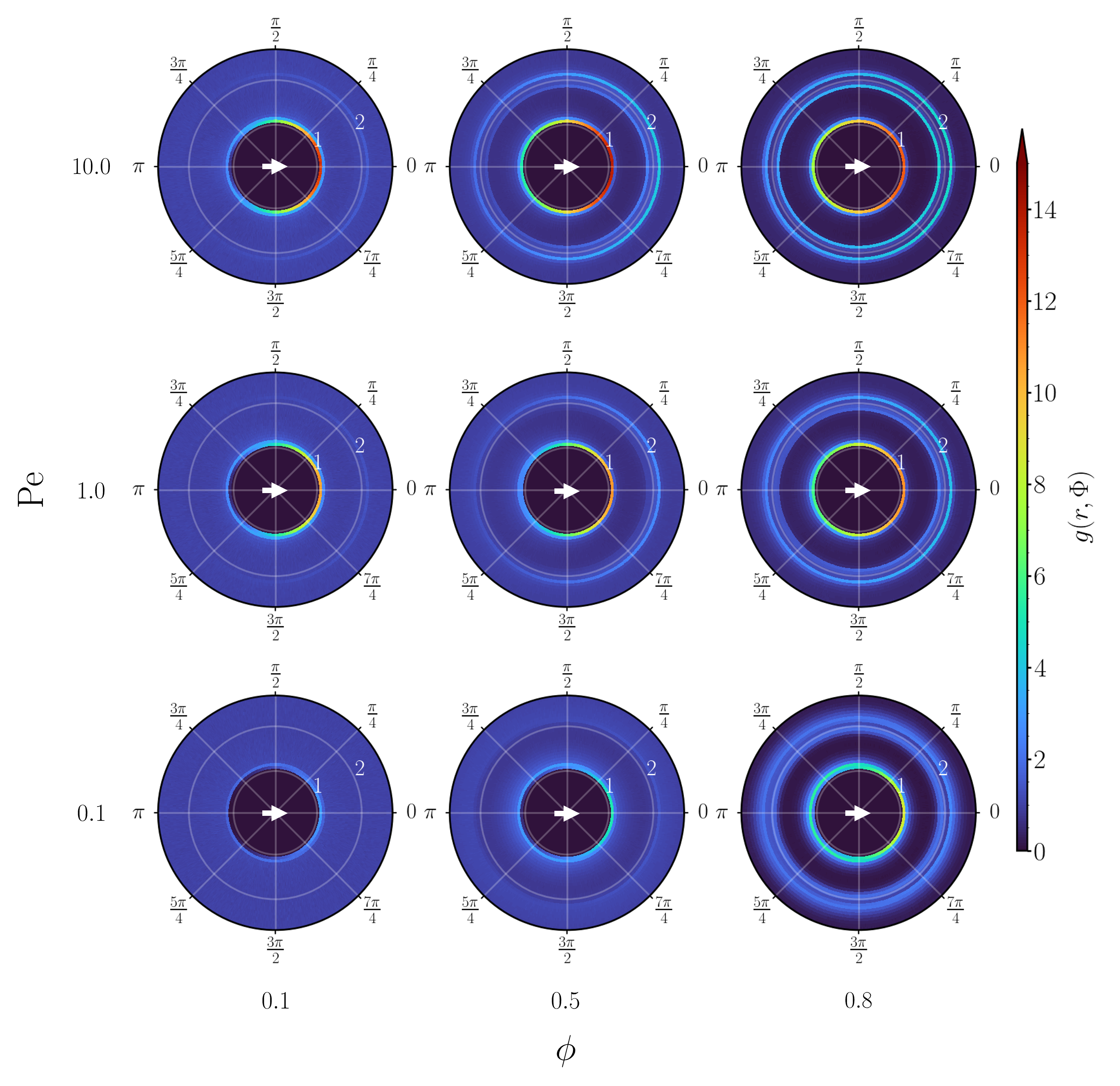}
\caption{\protect\small{The anisotropic pair correlation function \(g(r,\Phi)\) for a representative set of \(\mathrm{Pe}\) and \(\phi\).
The white arrows indicate the self-propulsion direction, defined by the orientational unit vector \(\hat{\mathbf q}\), and the distance \(r\) is normalized by the particle diameter \(\sigma\).
All measurements were performed using a system size of \(\mathcal{N}=100{,}000\).}}
\label{SM_Fig:1}
\end{figure}

\clearpage

\section{normalized force–force correlation functions}

\begin{figure}[h!]
\centering
\includegraphics[width=.7\textwidth]{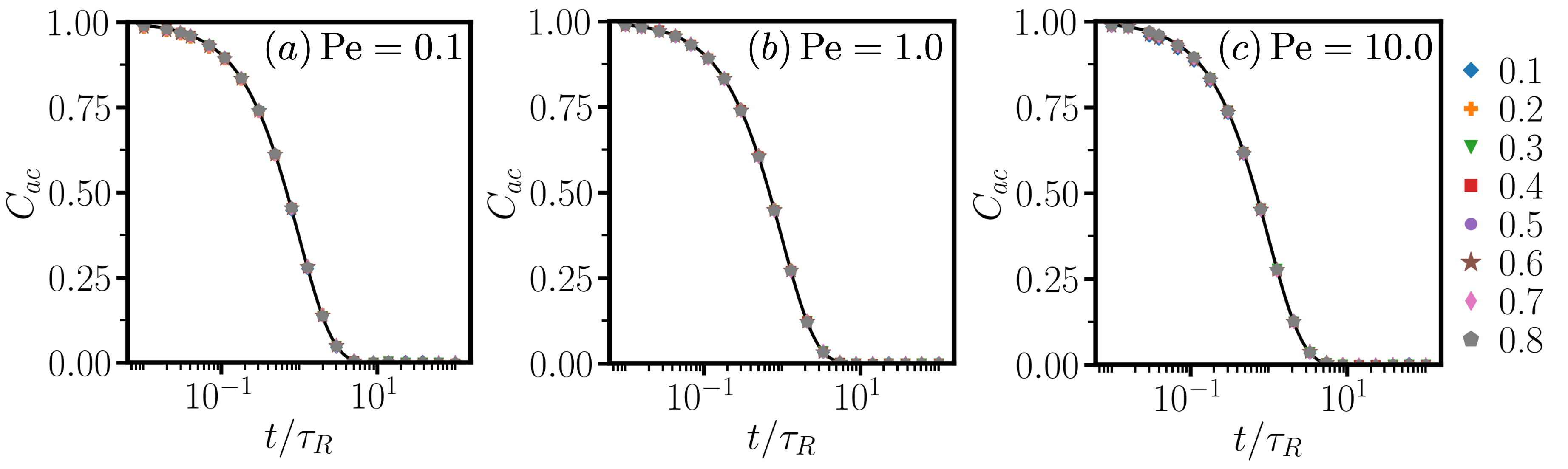}
\caption{\protect\small{Normalized force-force correlation \(C_{ac}(t)\) for a representative selection of \(\phi\) and \(\mathrm{Pe}\). The solid black lines denote the orientational autocorrelation $C_{aa}(t) =e^{-t/\tau_R}$.}}
\label{SM_Fig:2}
\end{figure}

\begin{figure}[h!]
\centering
\includegraphics[width=.7\textwidth]{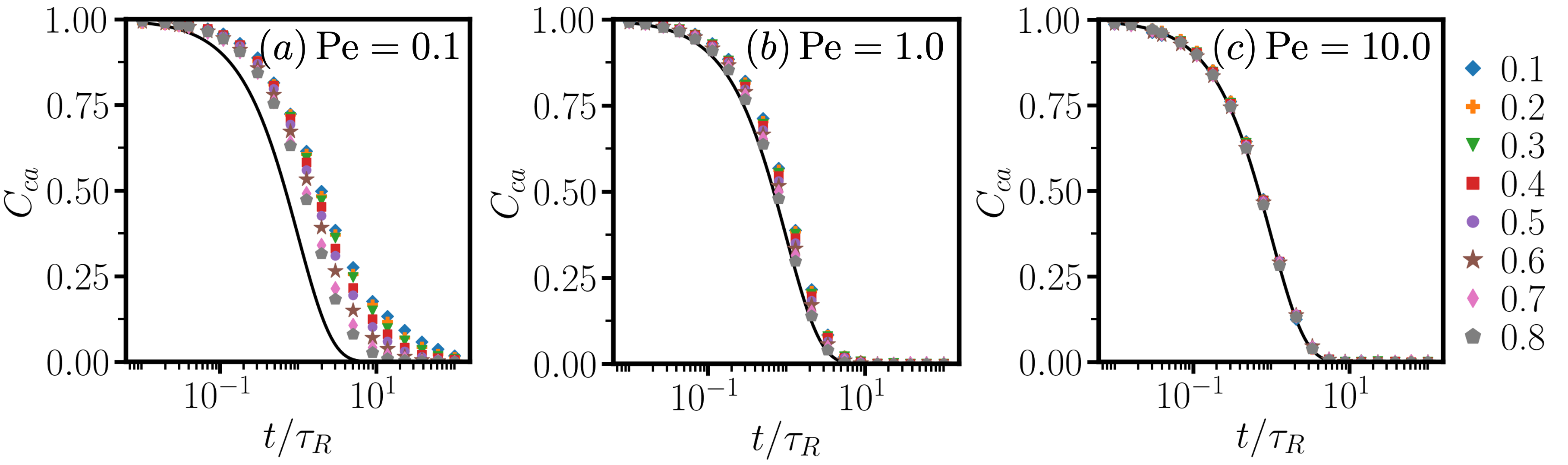}
\caption{\protect\small{Normalized force-force correlation \(C_{ca}(t)\) for a representative selection of \(\phi\) and \(\mathrm{Pe}\). The solid black lines denote the orientational autocorrelation $C_{aa}(t) =e^{-t/\tau_R}$.}}
\label{SM_Fig:3}
\end{figure}

\begin{figure}[h!]
\centering
\includegraphics[width=.7\textwidth]{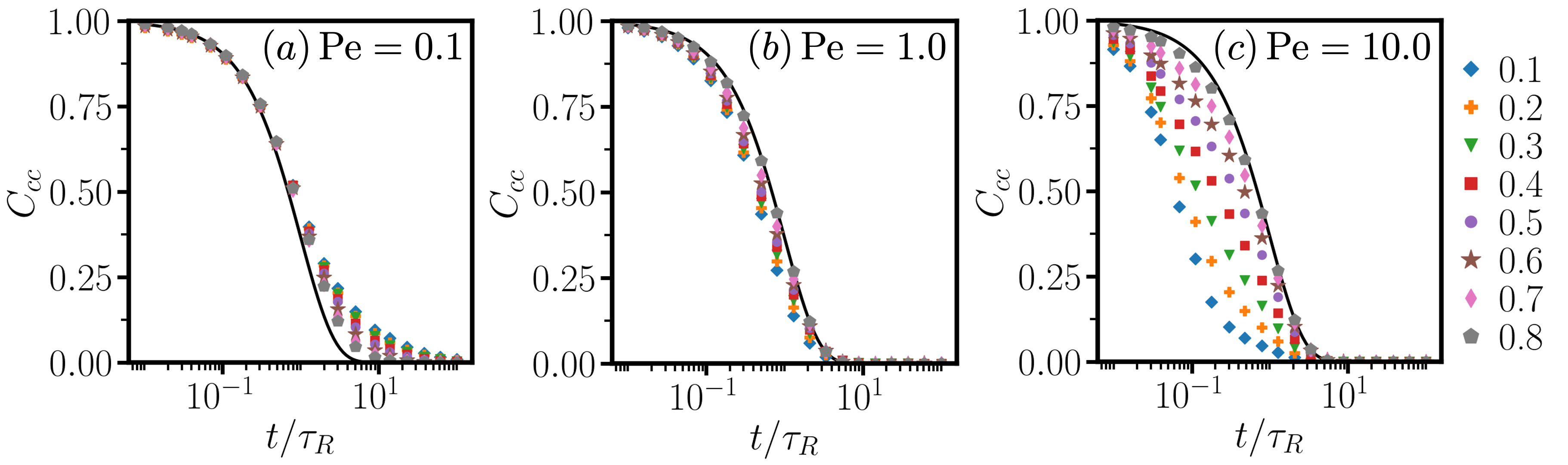}
\caption{\protect\small{Normalized force-force correlation \(C_{cc}(t)\) for a representative selection of \(\phi\) and \(\mathrm{Pe}\). The solid black lines denote the orientational autocorrelation $C_{aa}(t) =e^{-t/\tau_R}$.}}
\label{SM_Fig:4}
\end{figure}

\begin{figure}[h!]
\centering
\includegraphics[width=.7\textwidth]{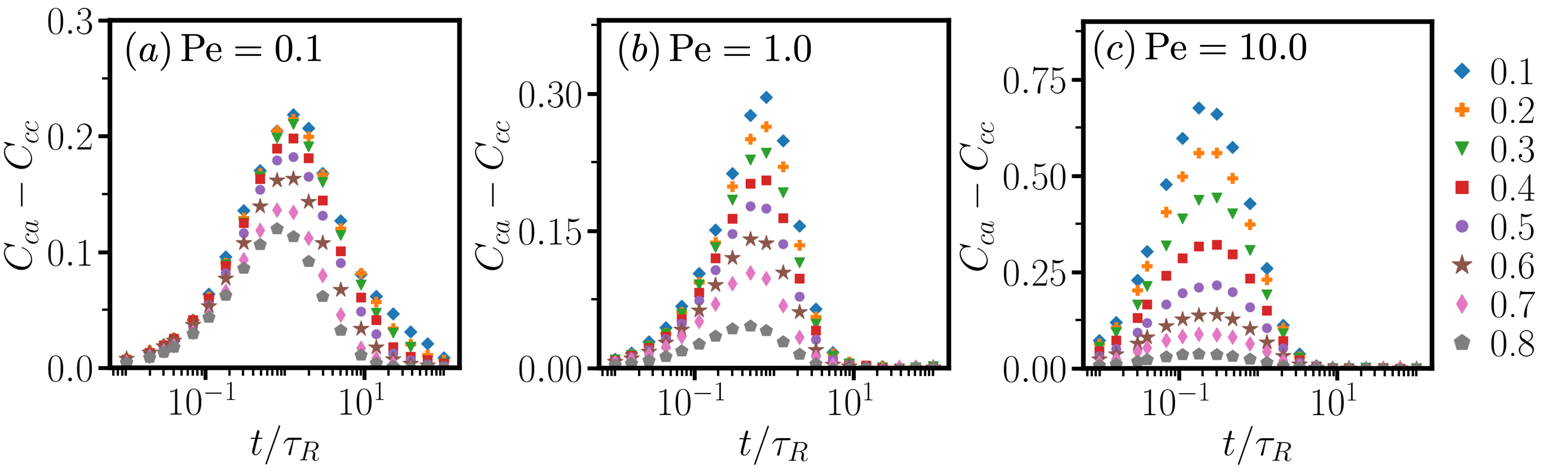}
\caption{\protect\small{Difference between normalized force-force correlations \(C_{ca}(t) - C_{cc}(t) \) for a representative selection of \(\phi\) and \(\mathrm{Pe}\).}}
\label{SM_Fig:5}
\end{figure}

\clearpage

\section{Large P\'eclet scaling of collisional force-orientation correlation time}

\noindent Here, we derive the large \(\mathrm{Pe}\) scaling for the collisional force-orientation correlation time \(\tau_{ca}\) and show that it approaches the intrinsic reorientation time \(\tau_R\).
By definition, the normalized force-force correlation function \(C_{ca}(\tau)\) is given by 
\begin{equation}
\label{eq:1}
C_{ca}(\tau)
\equiv
\frac{
\langle \mathbf F_c(\tau)\cdot \mathbf F_a(0)\rangle
}{
\langle \mathbf F_c(0)\cdot \mathbf F_a(0)\rangle
}
=
\frac{
\langle \mathbf F_c(\tau)\cdot \hat{\mathbf q}(0)\rangle
}{
\langle \mathbf F_c(0)\cdot \hat{\mathbf q}(0)\rangle
}
\, ,
\end{equation}
where \(\mathbf F_c\) is the collisional force and \(\mathbf F_a=\gamma U_a \hat{\mathbf q}\) is the active propulsion force.
We now decompose the collisional force into components parallel and perpendicular to the instantaneous propulsion direction,
\begin{equation}
\label{eq:2}
\mathbf F_c(t)
=
F_{\parallel}(t)\,\hat{\mathbf q}(t)
+
\mathbf F_{\perp}(t) \,,
\end{equation}
where \(F_{\parallel}(t)\equiv \mathbf F_c(t)\cdot \hat{\mathbf q}(t)\) and \( \mathbf F_{\perp}(t)\cdot \hat{\mathbf q}(t)=0\).
Substituting Eq.~(\ref{eq:2}) into Eq.~(\ref{eq:1}) gives
\begin{equation}
\label{eq:3}
C_{ca}(\tau)
=
\frac{
\langle F_{\parallel}(\tau)\,\hat{\mathbf q}(\tau)\cdot\hat{\mathbf q}(0)\rangle}{\langle  F_{\parallel}(0) \rangle}
+
\frac{\langle \mathbf F_{\perp}(\tau)\cdot\hat{\mathbf q}(0)\rangle
}{
\langle F_{\parallel}(0) \rangle
} \, ,
\end{equation}
where the denominator simplifies since, by definition, \(\mathbf F_{\perp}(0)\cdot\hat{\mathbf q}(0)=0\).
In a homogeneous and isotropic steady state, the second term of Eq.~(\ref{eq:3}) is expected to be negligible by symmetry, and we approximate \( C_{ca}(\tau) \) as  
\begin{equation}
\label{eq:4}
C_{ca}(\tau)
\approx
\frac{
\langle F_{\parallel}(\tau)\,\hat{\mathbf q}(\tau)\cdot\hat{\mathbf q}(0)\rangle
}{
\langle F_{\parallel}(0) \rangle
} \, .
\end{equation}
Next, we express the instantaneous parallel force \(F_{\parallel}(\tau)\) in terms of its steady-state average and fluctuation,
\begin{equation}
\label{eq:5}
F_{\parallel}(\tau)
=
\langle F_{\parallel}(\tau) \rangle
+
\delta F_{\parallel}(\tau) \, ,
\qquad
\langle \delta F_{\parallel}(\tau)\rangle = 0 \, .
\end{equation}
Substituting the parallel force decomposition in Eq.~(\ref{eq:5}) into Eq.~(\ref{eq:4}) and simplifying yields
\begin{equation}
\label{eq:6}
C_{ca}(\tau)
=
\langle \hat{\mathbf q}(\tau)\cdot\hat{\mathbf q}(0)\rangle
+
\frac{
\langle \delta F_{\parallel}(\tau)\,\hat{\mathbf q}(\tau)\cdot\hat{\mathbf q}(0)\rangle
}{
\langle F_{\parallel}(0) \rangle
}
=
C_{aa}(\tau)
+
\Delta C_{ca}(\tau) \, ,
\end{equation}
where \(C_{aa}(\tau)=\langle \hat{\mathbf q}(\tau)\cdot\hat{\mathbf q}(0)\rangle=e^{-\tau/\tau_R}\), and 
\(
\Delta C_{ca}(\tau)
\equiv
\langle \delta F_{\parallel}(\tau)\,\hat{\mathbf q}(\tau)\cdot\hat{\mathbf q}(0)\rangle/
\langle F_{\parallel}(0) \rangle
\)
captures additional correlations arising between force fluctuations and orientational persistence.
The collisional force-orientation correlation time is then defined as
\begin{equation}
\label{eq:32}
\tau_{ca}
=
\int_0^\infty
C_{ca}(\tau)\,d\tau 
=
\int_0^\infty
C_{aa}(\tau)\,d\tau
+
\int_0^\infty
\Delta C_{ca}(\tau)\,d\tau 
=
\tau_R
+
\tau_\Delta \, ,
\end{equation}
where \(\tau_\Delta=\int_0^\infty \Delta C_{ca}(\tau)\,d\tau\).
In the high-activity regime (\(\mathrm{Pe}>1\)), collisions occur at nearly fixed orientation, and local neighbor configurations renew on timescales short compared to \(\tau_R\).
In this limit, correlations between \(\delta F_{\parallel}\) and the orientational persistence decay rapidly, rendering \(\tau_\Delta\) small.
Consequently, \(\tau_{ca} \to \tau_R\) as \(\mathrm{Pe}\) increases, consistent with the asymptotic behavior observed in Fig.~\ref{SM_Fig:3} and Fig.~5(a) of the main text.

%\bibliography{SM}